# Bending modes metrology in the 14-15 μm region

M. Lamperti[1], R. Gotti[1], D. Gatti[1], M. K. Shakfa[2], E. Cané[3], F. Tamassia[3], P. Schunemann[4], P. Laporta[1], A. Farooq[2*], and M. Marangoni[1*]

[1]Dipartimento di Fisica - Politecnico di Milano and IFN-CNR, Via Gaetano Previati 1/C, 23900 Lecco, Italy

[2] King Abdullah University for Science and Technology, Clean Combustion Research Center, Thuwal 23955, Saudi Arabia

[3]Università di Bologna, Dipartimento di Chimica Industriale, Viale del Risorgimento 4, 40136 Bologna, Italy

[4]BAE Systems, Inc., MER15-1813, P.O. Box 868, Nashua, New Hampshire 03061-0868, USA

*aamir.farooq@kaust.edu.sa, marco.marangoni@polimi.it

**Frequency combs have triggered an impressive evolution of optical metrology across diverse regions of the electromagnetic spectrum, from the ultraviolet to the terahertz frequencies[1,2]. An unexplored territory, however, remains in the region of vibrational bending modes, mostly due to the lack of single-mode lasers in the long-wavelength (LW) part of the mid-infrared (MIR) spectrum[3]. We fill this gap through a purely MIR-based nonlinear laser source with tunability from 12.1 to 14.8 μm, optical power up to 110 μW, MHz-level linewidth and comb calibration. This enables the first example of bending modes metrology in this region, with the assessment of several $CO_2$-based frequency benchmarks with uncertainties down to 30 kHz, and the accurate study of the $\nu_{11}$ band of benzene, which is a significant testbed for the resolution of the spectrometer. These achievements pave the way for LW-MIR metrology[4], rotationally-resolved studies[5] and astronomic observations[6] of large molecules, such as aromatic hydrocarbons.**

The lack of continuously tunable single-mode lasers in the region from 13 to 20 μm has been a stumbling block for optical metrology and high-resolution spectroscopy in the region of vibrational bending modes[7]. Cryogenically cooled lead-salt diode lasers have almost disappeared in recent years[8]. They have been replaced by room-temperature distributed-feedback quantum-cascade-lasers (DFB-QCLs) that offer portability, stable operation and mode-hop-free tuning[9]. However, they are commercially available only below 13 μm[10], whereas the emerging technology of InAs-based long-wavelength QCLs is still under development and not available for applications[11]. An alternative approach is difference frequency generation (DFG) from continuous-wave (cw) Ti:sapphire[12,13] or diode[14] lasers, but this is accompanied by extremely low optical powers (10-100 nW range) that impair the acquisition of absorption spectra at high signal-to-noise ratio (SNR). The limitations of laser technology have established Fourier-Transform spectroscopy driven by incoherent light sources as a gold standard for the LW-MIR region, at the price of low resolution (0.0007 cm$^{-1}$ in the best cases[15]) and of no absolute calibration for the frequency axis. A powerful solution to both issues is direct comb spectroscopy[16], which has recently conquered the LW-MIR region by a dual-comb approach[17] that enabled snapshots of entire bands from 6.7 to 16.7 μm at high temporal and spectral resolution[18]. This achievement does not remove, however, the need for cw probe lasers to observe single lines with high accuracy [4,7,19] and extreme speed[20] in a region where the Doppler width and the typical line densities are consistently below the spacing between adjacent comb modes. Furthermore, to the best of our knowledge, optical metrology has not been demonstrated so far by direct comb spectroscopy in the LW-MIR region.

Our spectrometer is schematically shown in Fig. 1a. The laser source is based on the DFG process between two MIR sources, namely a cw DFB-QCL and a $CO_2$ laser, in an orientation-patterned gallium arsenide (OP-GaAs) crystal[21]. The DFG laser provides highly coherent radiation between 12.1 and 14.8 μm with optical power up to 110 μW and a 2.2 MHz linewidth. The large spectral range comes from the coarse wavelength adjustment of the $CO_2$ laser (from 9.23 to 10.86 μm) combined with the fine temperature-based tuning of the QCL (over ~4 cm$^{-1}$ around 5.69 μm) and the fan-out structure of the poling periods of the OP-GaAs crystal (see Methods for details). The optical power is 2-3 orders of magnitude better than previous DFG sources starting from cw Ti:sapphire lasers with GaSe crystals[12,13]. The obtained power agrees within 15% of our calculations using 63 pm/V as the effective nonlinear coefficient. Such an excellent agreement reflects the quality of the crystal and a very favourable MIR-MIR interaction that ensures similar diffraction conditions for the interacting beams and thus optimal spatial overlap over a long crystal. The absolute frequency calibration of the DFG radiation comes from the referencing of both QCL and $CO_2$ lasers to a 100 MHz comb *via* sum frequency generation[22] (see Methods for details). As sketched in Fig. 1b, the $CO_2$ laser remains offset-locked to the nearest comb mode during the measurements, whereas the frequency of the temperature-tuned DFB-QCL is tracked against the comb by real-time fast Fourier transform (FFT) analysis and barycentre calculation of its beat note. Synchronous acquisition of the beat note and gas transmission allows spectra to be straightforwardly calibrated.

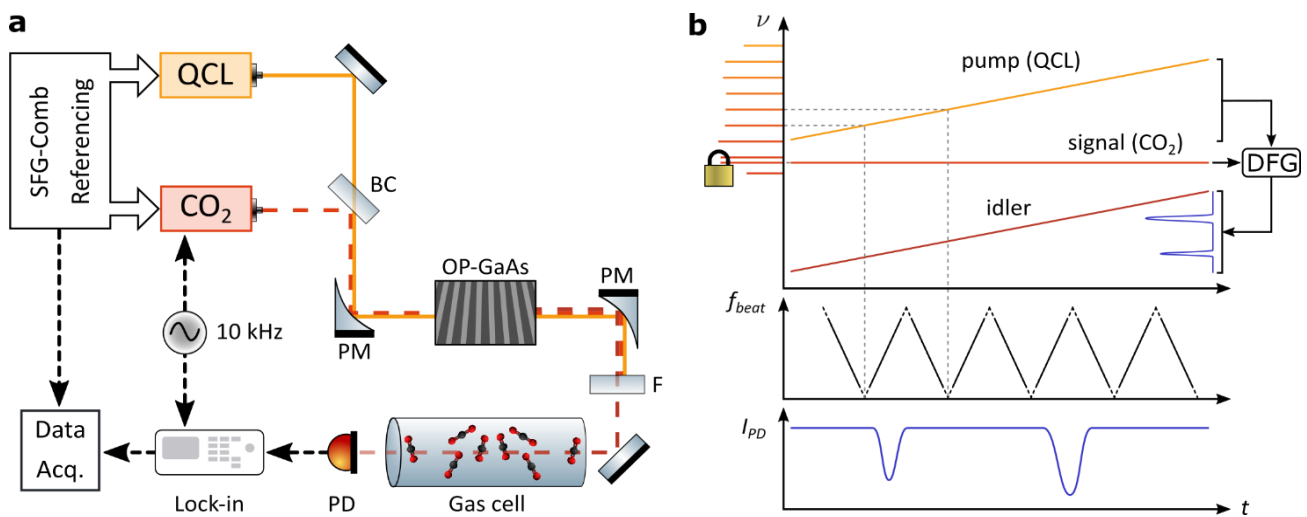

**Fig. 1 | Comb-referenced nonlinear laser source for the LW-MIR. a.** Experimental lay-out. BC: beam combiner; PM: parabolic mirror; F: low-pass filter; PD: liquid-nitrogen cooled photodiode; SFG: sum frequency generation. **b.** Pictorial view of the DFG process between the comb-locked $CO_2$ laser and the QCL. The synchronous acquisition of the QCL-comb beat note ($f_{beat}$) and of the idler absorption ($I_{PD}$) allow absolute calibration of the spectra over the measurement time. For graphical simplicity pump and signal are shown to beat directly with a MIR comb, but in practice this happens through the interposition of a sum frequency generation process (see Methods for details).

Figure 2a shows an example of comb-calibrated transmission spectrum for some intense lines of the $\nu_2$ band of $CO_2$ in the 675-689 cm$^{-1}$ (14.5-14.8 μm) range. Lines are well isolated and in excellent qualitative agreement with the HITRAN[23] simulations.

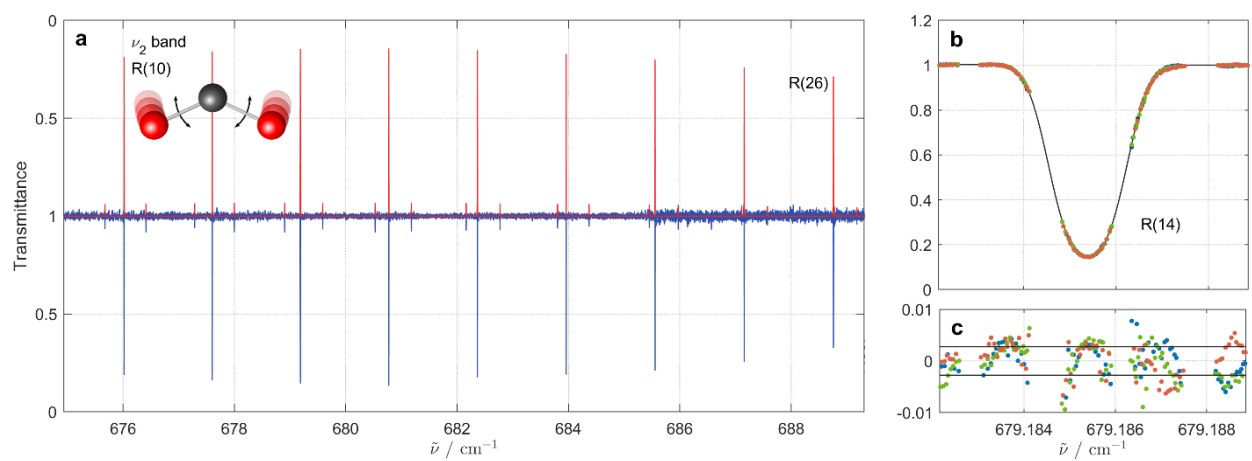

**Fig. 2 | Carbon dioxide spectroscopy. a.** Absorption spectrum of the $\nu_2$ band of $CO_2$ (blue) as compared to HITRAN simulations (red). Experimental conditions: interaction length = 0.67 m, pressure = 0.93 Pa, room temperature. The scattering of baseline points is due to etalons that have been fitted out in the proximity of lines. **b.** Zoomed-in view of the R(14) line, with independent measurements of the same line (coloured dots), and fit with a Voigt profile (black line). Each measurement is composed of 170 points acquired over 2 seconds. **c.** Residuals from the Voigt fitting of each individual measurement, showing the presence of a periodic structure due to parasitic etalons in the setup. The black lines delimit the region of ± 1 standard deviation of the residuals (0.3 % of the line amplitude).

Figure 2b shows three independent experimental traces of the R(14) line, reported in different colors to highlight their reproducibility over the two measurement axes. The presence of spectral gaps is due to the missed calibration that occurs when the comb-QCL beat note approaches either DC or half the spacing between neighboring comb modes (50 MHz, see Methods for details). With an additional spectral acquisition at a slightly different comb repetition rate, the gaps would disappear. They do not, however, represent an issue when fitting the data (black line in Fig. 2b): the residuals in Fig. 2c refer to a Voigt model and give an SNR of approximately 300 with a measurement time of 2 seconds. This translates to a relatively small statistical uncertainty of about 140 kHz on the inferred line center frequency, mostly due to the reduced Doppler broadening (30 MHz level) given by the LW-MIR range and to the negligible collisional broadening at the low pressure used here.

For each measured line, Fig. 3 shows the difference between its absolute line-centre frequency and the value in the HITRAN database, together with an error bar dominated by statistical uncertainties (see Methods for details). Some lines have been intentionally measured several times to improve their line-centre uncertainty and to provide frequency standards in a region where absolute calibration has not been reported so far, to the best of our knowledge. The weighted deviation from HITRAN, averaged over 24 transitions, amounts to a remarkably low 0.34 MHz, which is consistent with the declared HITRAN uncertainty (0.3-3 MHz). The line centres reported for each line in the Supplementary Table 1 offer a metrological benchmark to calibrate past and future spectra in the 670-720 cm$^{-1}$ (14-15 μm) region.

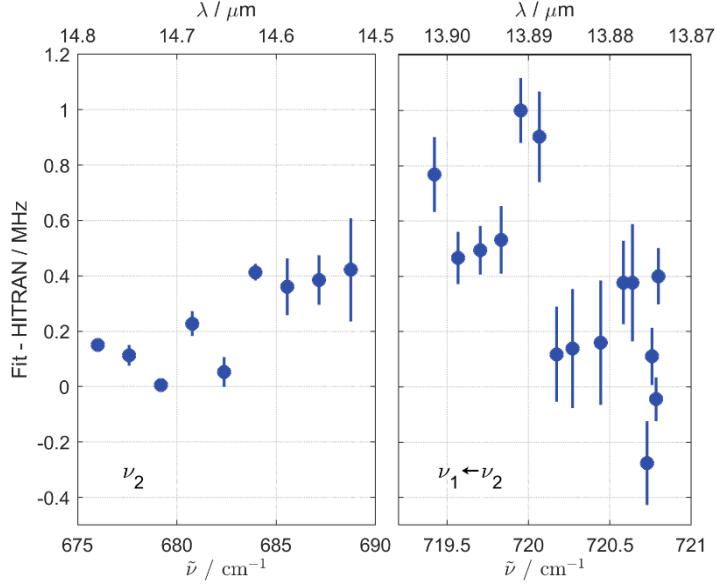

**Fig. 3 | Carbon dioxide: experimental line centres vs HITRAN values.** Difference between experimental and HITRAN line centres for several lines of the R branch of the $\nu_2$ band (left panel) and of the Q branch of the $\nu_1 \leftarrow \nu_2$ band (right panel). The variation of error bars comes from different number of measurements per line.

In Fig. 4, we report the experimental spectrum of the R branch of the $\nu_{11}$ band of benzene over 675-688 cm$^{-1}$. This intense, parallel band is an interesting testbed for the spectral resolution and frequency repeatability of the spectrometer because of the large density of rotational lines. Apart from transitions with quantum number $K$ = 0, 1 and 2, whose separation is smaller than the Doppler width, all $K$ lines are well resolved. This is shown in the insets in Fig. 4 and is particularly evident at large $J$. Lines from three hot bands, namely $\nu_{11} + \nu_{16} - \nu_{16}$, $\nu_{11} + \nu_6 - \nu_6$ and $2\nu_{11} - \nu_{11}$, are also present in the recorded spectral region, but the analysis was restricted to the 729 ro-vibration transitions of the $\nu_{11}$ band, with $J$ up to 38 and $K$ up to 36, since these are mostly isolated, medium intensity lines. We adopted a weighted least-squares fitting procedure to determine highly accurate term values of the excited ro-vibration levels, according to the equation:

$$E_v(J,k) = E_v^0 + B_v[J(J+1) - k^2] + C_v k^2 - D_{v,J}[J(J+1)]^2 - D_{v,JK}J(J+1)k^2 - D_{v,K}k^4 \\ + H_{v,J}[J(J+1)]^3 + H_{v,JK}[J(J+1)]^2 k^2 + H_{v,KJ}[J(J+1)]k^4 + H_{v,K}k^6 \quad (1)$$

where v stands for $\nu_{11}$ = 1 and $k = \pm K$. The ground state term values were derived from Ref. 24 and Eq. (1) with v = 0, $E_v^0$ = 0 and keeping all $H$ centrifugal distortion parameters fixed to zero. A unitary weight was attributed to the isolated lines, whose wavenumber precision is estimated to be 1 x 10$^{-5}$ cm$^{-1}$. The weights of blended or weak lines were reduced to 0.01, and, in addition, for lines with multiple assignments, the weight was further reduced by a factor equal to the number of overlapped transitions. The parameters obtained from the best fit are listed in Table 1, along with those of the ground state for comparison.

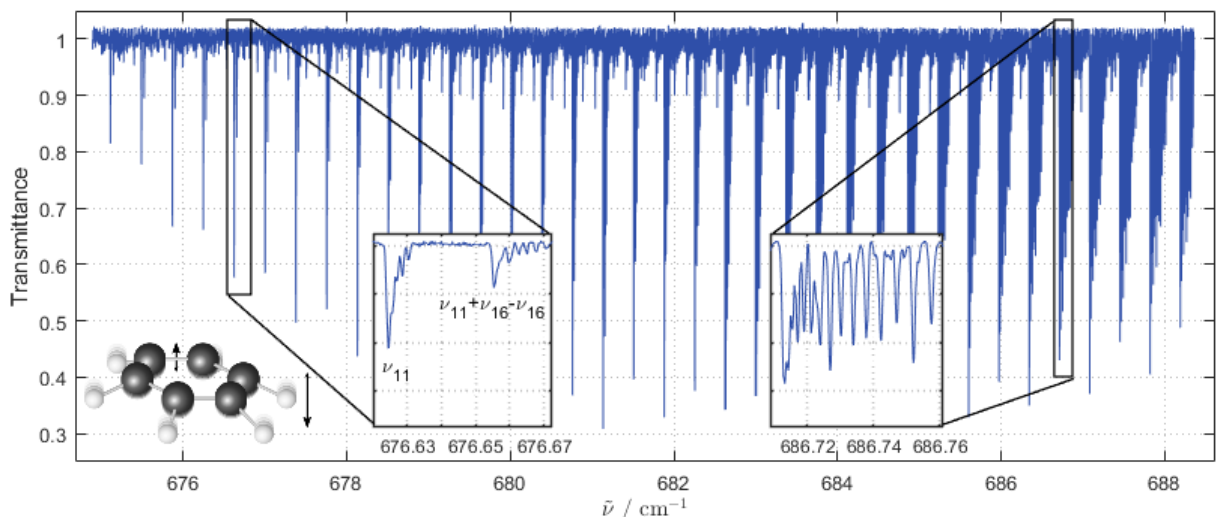

**Fig. 4 | Absorption spectrum of the R-branch of the $\nu_{11}$ band of benzene.** Experimentally measured transmission at the output of a 0.67 m long cell at a pressure of 13.3 Pa and room temperature. The spectral features have a typical shape, with the *K*-structure degrading regularly towards higher wavenumbers and an apparent intensity alternation of subsequent lines. The insets show zoomed-in views of $R_K(6)$ and $R_K(33)$ features (left and right inset, respectively): for the latter, the *K*-splitting is much more manifest together with intensity alternation of *K* components, owing to spin statistics, that varies as 10:11:9:14, for $K = 6p$, $6p\pm1$, $6p\pm2$, $6p\pm3$, respectively, ($p = 0, 1, 2, 3, ...$ )[25].

All parameters reported in Table 1 are statistically well determined. The uncertainty of the band origin ($E_v^0$) has been reduced by about one third with respect to its value in Ref. 26. The *B*, *C*, $D_J$, $D_{JK}$, and $D_K$ constants are consistent and close to those of the ground state[24]. The values of *B*, $D_J$, and $D_{JK}$ are more accurate than those in Ref. 26 by one order of magnitude. The values of *C*, $D_K$, and $H_J$ constants of $v_{11} = 1$ are determined here for the first time. The standard deviation of the fit, $5.1 \times 10^{-5}$ cm$^{-1}$ (1.5 MHz), is one order of magnitude smaller than in the past literature for the same band[26]. At the same time, it is about 10 times higher than for single $CO_2$ lines due to a combination of factors, namely, the many partially overlapped lines and the difficulty to access intensity baseline due to the congested spectrum. The absolute calibration of the frequency axis allows, for the first time, the stacking of repeated measurements and the determination of absolute line positions at these wavelengths.

**Table 1.** Spectroscopic parameters (cm$^{-1}$) for the ground state and for the $v_{11} = 1$ excited state of benzene [a]

|   |   | Ground state | $v_{11} = 1$ |
|---|---|---|---|
| $E_v^0$ |  |  | 673.9751463(121) |
| B |  | 0.1897727 | 0.1896373408(614) |
| C |  | 0.0948863 | 0.0949217259(632) |
| $D_J$ | $\times 10^8$ | 3.34 | 3.41413(847) |
| $D_{JK}$ | $\times 10^8$ | –6.57 | -6.54650(372) |
| $D_{JK}$ | $\times 10^8$ | 3.26 | 3.23876(301) |
| $H_J$ | $\times 10^{12}$ | 0.0 | 0.1846(343) |
| No. of data |  |  | 729 |
| σ (fit) | $\times 10^5$ |  | 5.1 |

[a] Standard uncertainties (1σ) in parentheses refer to the least significant digits.
[b] Ref. 24.

Overall, the metrological approach presented here considerably expands the portfolio of laser tools to probe molecules in the gas phase with high resolution and accuracy. This is proactive for the study of large molecules through the unique fingerprint features offered by the LW-MIR spectral region, for example molecules of the BTEX family (benzene, toluene, ethylbenzene, xylenes) that are of high interest for environmental modelling applications. In the case of benzene, which lacks a permanent dipole moment and, therefore, cannot be detected by pure rotational spectroscopy, infrared bands such as the $\nu_{11}$ explored here represent the clue to assess their presence in remote spatial environments, such as planetary atmospheres[6,27,28]. On another front, by the use of a sub-hertz linewidth comb and phase-locking loops for pump and signal lasers, the spectrometer could reach the level of stability and spectral purity needed for tests of fundamental physics on molecular samples[29,30] in a spectral region not explored so far.

## METHODS

*Difference Frequency Generation*

The pump radiation in the DFG process is provided by a distributed-feedback quantum-cascade-laser (DFB-QCL, from Alpes Lasers) emitting around 5.69 μm. Tuning its temperature from 20 °C to -10 °C results in a continuous spectral coverage from 1757 to 1761 cm$^{-1}$ (i.e., over 4 cm$^{-1}$) with an optical power from 17 to 48 mW, respectively. The signal radiation is generated by a $CO_2$ laser (L20GD, Access Laser) operating in a pulse modulation mode ("super-pulse mode") with a duty cycle of 20 % and a modulation frequency of 10 kHz. Its emission can be finely piezo-tuned over few tens of MHz and coarsely tuned by grating over ~ 80 lines, from 9.23 to 10.71 μm, with peak powers varying from 3 to 27 W depending on the line. The wavelength combination of the two lasers provides access to 12.1-14.9 μm wavelength range for the idler beam, apart from 10 cm$^{-1}$ gaps at 12.6 and 14.4 μm and a 30 cm$^{-1}$ gap at 13.3 μm due to the discontinuities between adjacent $CO_2$ lasing bands. The OP-GaAs crystal used for DFG is 35 mm long and equipped with poling periods from 183 to 203 μm in a fan-out structure, which ensures phase matching over the entire tuning range available for the nonlinear laser source (upper bound at 14.8 μm rather than 14.9 μm). The width of the crystal is 20 mm, which is sufficiently large to discard any efficiency drop due to non-uniform phase-matching conditions across the beams. In fact, the poling period changes by less than 0.2 μm, which is far below the phase matching bandwidth of 0.7 μm, over beam diameters of 120 and 200 μm for pump and signal, respectively. As expressed in terms of wavenumbers, the phase-matching bandwidth amounts to 4.2 cm$^{-1}$, which implies that there is no need for adjustment of the crystal position while tuning the QCL. The optical power generated by the DFG considerably depends on both the QCL temperature and the $CO_2$ lasing line, but it remains above 20 μW (peak power) if the $CO_2$ emission is chosen among the 70 % most intense lines, as is the case for all measurements shown here.

**Comb referencing and spectra acquisition**

Pump (5.7 μm) and signal (9.2-10.7 μm) lasers are referenced to the same 1.9 μm Tm:fiber frequency comb through two independent sum frequency generation (SFG) processes, in $AgGaSe_2$ and ZnGaP (ZGP) crystals, respectively. The SFG processes shift the comb frequency $\nu_n = f_{ceo} + nf_{rep}$ ($f_{ceo}$ and $f_{rep}$ being carrier-envelope-offset and repetition frequency of the comb, respectively) by the pump ($\nu_p$) and signal ($\nu_s$) frequency, thereby generating two replicas of the comb around 1.4 ($\nu_{n,p} = \nu_n + \nu_p$) and 1.6 μm ($\nu_{n,s} = \nu_n + \nu_s$) for pump and signal, respectively. When these replicas are superimposed with a coherent continuum ($\nu_m = f_{ceo} + mf_{rep}$) generated from the original comb and extending down to 1.4 μm, a beat note is extracted, $f_{beat} = \pm(\nu_{n,p/s} - \nu_m) = \pm(\nu_{s,p} - (m-n)f_{rep})$ that straightforwardly links the unknown pump/signal frequency to an integer number of $f_{rep}$, independently of $f_{ceo}$. In our case, $f_{rep}$ is stabilized to a GPS-tracked low-noise radiofrequency (RF) signal at 100 MHz while $f_{ceo}$ is left free running. From an analysis of their beat notes (Supplementary Fig. 1 and Fig. 2), QCL and $CO_2$ lasers give an almost Gaussian instrumental line shape of 1.6 and 1.5 MHz, respectively, which sums up to 2.2 MHz. The relatively large contribution from the $CO_2$ laser comes from the adopted super-pulse mode and it could be suppressed by choosing a cw regime followed by external modulation, at the expense of DFG power. The $CO_2$ beat note is locked to an RF local oscillator via piezo-feedback to the grating position, whereas the QCL beat note is measured in real time by a 100 MS/s DAQ, followed by an FFT conversion at every 1024 samples and a barycentre calculation via FPGA (PXIe-7961 FPGA board and NI-5781 add-on, National Instruments). The same board digitizes the lock-in output in order to synchronize horizontal and vertical axis of the measurement. An example of raw data for QCL beat note and gas absorption is given in the Supplementary Fig. 3. The acquisition of a 4 $cm^{-1}$ large spectrum is typically accomplished in 12 minutes (tuning rate of 180 MHz/s) to prevent laser frequency changes by more than its linewidth over the integration time of 10 ms set for the lock-in. A higher speed would be possible with a faster modulation of the $CO_2$ laser and a correspondingly smaller integration time. Spectra larger than 4 $cm^{-1}$, as those in Fig. 2 and Fig. 4, are acquired piecewise, due to the need of changing the $CO_2$ laser frequency and correspondingly adjusting the phase-matching condition for DFG. No tuning of the phase-matching angle is necessary for the SFG processes because of the much larger phase-matching bandwidths given by the shorter interaction length (6 mm) and the type of phase matching (birefringence rather than quasi-phase-matching). The assignment of the comb mode order for correct calibration of the frequency axis is straightforward because of the low uncertainty of both $CO_2$ and $C_6H_6$ lines, well below the comb mode spacing (100 MHz).

**Uncertainty budget**

A 240 MHz-large spectrum of a single isolated line, such as that shown in Fig. 2b, is typically acquired in 2 s and encompasses 170 spectral points: these are almost evenly distributed in 5 groups separated at every 50 MHz, namely at every half $f_{rep}$, by 18 MHz blanks due to the missed tracking of the comb-QCL beat note. In these conditions, the absorption noise ($3·10^{-3}$ on the most intense lines) and the frequency uncertainty on each spectral point (29 kHz) account for about 70% and 5%, respectively, of the observed 140 kHz root mean square (rms) deviation of the line centre for repeated measurements. The missing contribution comes from distortions of the spectral baseline that change from measurement to measurement and are not properly fit out. The statistical uncertainty dominates over systematic effects, due to the negligible impact from the instability of the GPS-based frequency standard (2 kHz over 2 s), the inaccuracy of the absolute pressure gauge and of the pressure shift coefficients used to extrapolate line centres to zero pressure (< 2 kHz, thanks to the low pressure), and the inappropriateness of the Voigt fitting model (< 3 kHz, due to the negligible asymmetry of line profiles in the low-pressure collisional regime).

**ACKNOWLEDGMENTS**

The authors acknowledge a financial contribution from the cooperative project OSR-2019-CCF-1975.34 between Politecnico di Milano and King Abdullah University of Science and Technology (KAUST) and by the project EMPATIA@Lecco ID: 2016-1428. F.T. and E.C. thank the University of Bologna for RFO funds.

**AUTHOR CONTRIBUTIONS**

M.M. and A.F. conceived the experiments. M.L and K.S. realized and characterized the DFG source while D.G. and R.G. took care of its referencing to the comb. M.L., R.G., D.G. performed the spectroscopic measurements. F.T., E.C. took care of fitting and interpreting benzene spectra, while M.L., R.G. and P.L. of carbon dioxide spectra. P.S. designed and provided the nonlinear crystal. M.M. took care of the first draft writing. All authors contributed to and edited the manuscript.

# SUPPLEMENTARY MATERIAL

# Bending modes metrology in the 14-15 μm region

M. Lamperti[1], R. Gotti[1], D. Gatti[1], M. K. Shakfa[2], E. Cané[3], F. Tamassia[3], P. Schunemann[4], P. Laporta[1], A. Farooq[2*], and M. Marangoni[1*]

[1]Dipartimento di Fisica - Politecnico di Milano and IFN-CNR, Via Gaetano Previati 1/C, 23900 Lecco, Italy
[2] King Abdullah University for Science and Technology, Clean Combustion Research Center, Thuwal 23955, Saudi Arabia
[3]Università di Bologna, Dipartimento di Chimica Industriale, Viale del Risorgimento 4, 40136 Bologna, Italy
[4]BAE Systems, Inc., MER15-1813, P.O. Box 868, Nashua, New Hampshire 03061-0868, USA
*aamir.farooq@kaust.edu.sa, marco.marangoni@polimi.it

**Characterization of the DFG linewidth**

The linewidth of the DFG laser source has been estimated from the measured beat notes of the QCL and $CO_2$ lasers with the comb (Fig. S1 and S2, respectively). These have an almost Gaussian-like shape, which justifies a quadrature sum of their widths to have a reliable estimation of the DFG linewidth. The contribution of the comb linewidth (~ 100 kHz) to the beat notes (~ 1.5 MHz) is negligible.

For both lasers, the beat note has been measured via the 100 MS/s DAQ used for spectral acquisition, by subdividing the beat note samples in groups of 1024 points and calculating the barycentre of the electrical spectrum for each group via FFT. This is equivalent to repeatedly measuring the average laser frequency (as compared to the comb) over a measurement time of 10 μs (1024 points at every 10 ns). The distribution of these frequencies is shown in Fig. S1 and S2 for QCL and CO2 laser, respectively. The width of these distributions provides an estimation value for their linewidths.

**Supplementary Fig. 1**

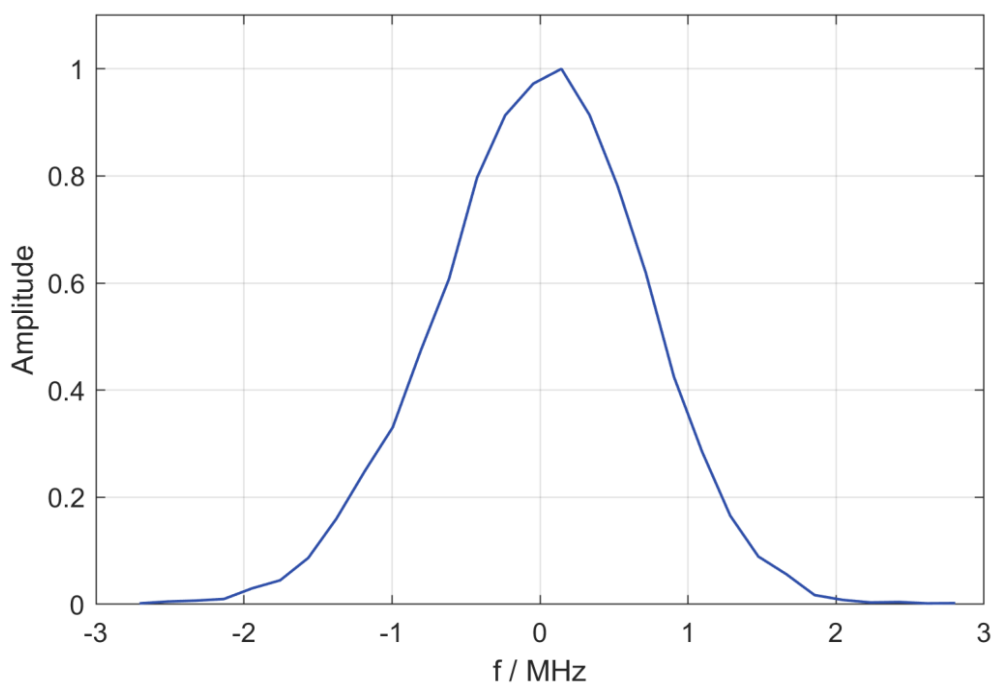

**Fig. S1 | Beat note spectrum between QCL and comb.** It is obtained as a histogram of the barycentre frequencies measured by the acquisition board over subsequent time windows of 10 μs.

**Supplementary Fig. 2**

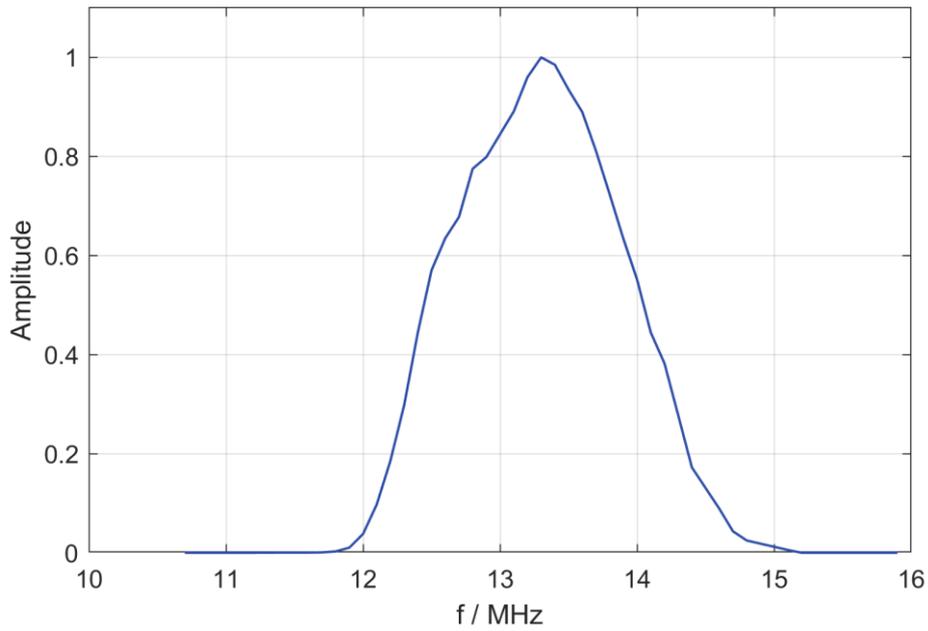

**Fig. S2 | Beat note spectrum between CO₂ laser and comb.** It is obtained as a histogram of the barycentre frequencies measured by the acquisition board over subsequent time windows of 10 µs. The slight asymmetry derives from the pulsed operation of the laser.

**Supplementary Fig. 3**

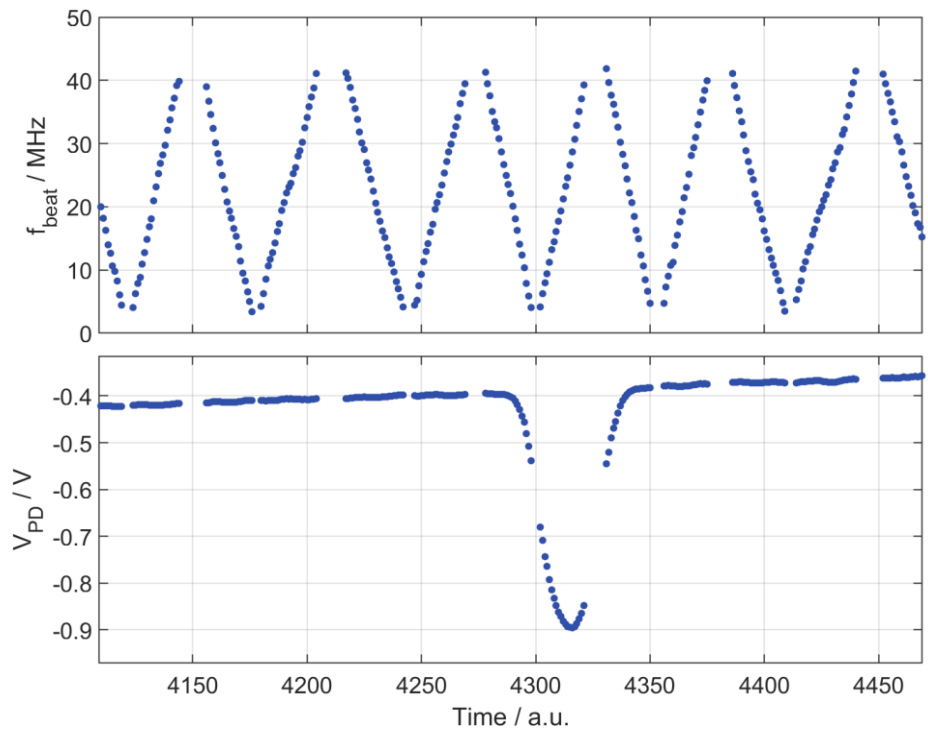

**Fig. S3 | Raw measurement data for a CO₂ line.** Time behaviour of the beat note frequency between QCL and comb (upper panel) and of the photodetector signal (lower panel).

**Supplementary Table 1**

**Table S1 | Carbon dioxide line list.**

Observed and calculated wavenumbers of the transitions assigned to $\nu_2$ and $\nu_1 \leftarrow \nu_2$ for $CO_2$, ordered for increasing energy.

| Band | Line | Obs / cm$^{-1}$ | HITRAN / cm$^{-1}$ | Obs-HITRAN / MHz |
|---|---|---|---|---|
| $\nu_2$ | R(10) | 676.01943603(66) | 676.0194310 | 0.151 |
| $\nu_2$ | R(12) | 677.6008247(13) | 677.6008210 | 0.113 |
| $\nu_2$ | R(14) | 679.18538721(70) | 679.1853870 | 0.006 |
| $\nu_2$ | R(16) | 680.7731066(15) | 680.7730990 | 0.228 |
| $\nu_2$ | R(18) | 682.3639287(18) | 682.3639270 | 0.052 |
| $\nu_2$ | R(20) | 683.9578497(10) | 683.9578360 | 0.413 |
| $\nu_2$ | R(22) | 685.5548060(34) | 685.5547940 | 0.360 |
| $\nu_2$ | R(24) | 687.1547798(30) | 687.1547670 | 0.386 |
| $\nu_2$ | R(26) | 688.7577330(62) | 688.7577190 | 0.422 |
| $\nu_1 \leftarrow \nu_2$ | Q(36) | 719.4226506(45) | 719.4226250 | 0.769 |
| $\nu_1 \leftarrow \nu_2$ | Q(34) | 719.5662865(32) | 719.5662710 | 0.466 |
| $\nu_1 \leftarrow \nu_2$ | Q(32) | 719.7026444(29) | 719.7026280 | 0.494 |
| $\nu_1 \leftarrow \nu_2$ | Q(30) | 719.8315747(41) | 719.8315570 | 0.532 |
| $\nu_1 \leftarrow \nu_2$ | Q(28) | 719.9529553(39) | 719.9529220 | 1.000 |
| $\nu_1 \leftarrow \nu_2$ | Q(26) | 720.0666311(55) | 720.0666010 | 0.905 |
| $\nu_1 \leftarrow \nu_2$ | Q(24) | 720.1724819(58) | 720.1724780 | 0.118 |
| $\nu_1 \leftarrow \nu_2$ | Q(22) | 720.2704506(72) | 720.2704460 | 0.139 |
| $\nu_1 \leftarrow \nu_2$ | Q(18) | 720.4422813(75) | 720.4422760 | 0.160 |
| $\nu_1 \leftarrow \nu_2$ | Q(14) | 720.5814236(50) | 720.5814110 | 0.378 |
| $\nu_1 \leftarrow \nu_2$ | Q(12) | 720.6385565(71) | 720.6385440 | 0.377 |
| $\nu_1 \leftarrow \nu_2$ | Q(8) | 720.7276548(51) | 720.7276640 | -0.276 |
| $\nu_1 \leftarrow \nu_2$ | Q(6) | 720.7595686(35) | 720.7595650 | 0.110 |
| $\nu_1 \leftarrow \nu_2$ | Q(4) | 720.7829825(26) | 720.7829840 | -0.045 |
| $\nu_1 \leftarrow \nu_2$ | Q(2) | 720.7979103(34) | 720.7978970 | 0.401 |

**Supplementary Table 2**

**Table S2 | Benzene line list.**

Observed and calculated wavenumbers of the transitions assigned to $\nu_{11}$ for benzene, ordered for increasing energy.
The headings J', K' and J", K" indicate the values of the rotational quantum numbers for the upper and lower states of transitions, respectively. The weights used in the fitting of band constants are also reported (see the main text for description).

| J' | K' | J" | K" | OBS / cm$^{-1}$ | CALC / cm$^{-1}$ | OBS-CALC / cm$^{-1}$ | WEIGHT |
|---|---|---|---|---|---|---|---|
| 3 | 0 | 2 | 0 | 675.11231800 | 675.11215451 | 0.00016349 | 0.0033 |
| 3 | 1 | 2 | 1 | 675.11231800 | 675.11232569 | -0.00000769 | 0.0033 |
| 3 | 2 | 2 | 2 | 675.11231800 | 675.11283922 | -0.00052122 | 0.0033 |
| 4 | 0 | 3 | 0 | 675.49087000 | 675.49061191 | 0.00025809 | 0.0033 |
| 4 | 2 | 3 | 2 | 675.49087000 | 675.49129713 | -0.00042713 | 0.0033 |
| 4 | 1 | 3 | 1 | 675.49087000 | 675.49078321 | 0.00008679 | 0.0033 |
| 4 | 3 | 3 | 3 | 675.49209300 | 675.49215368 | -0.00006068 | 0.0100 |
| 5 | 0 | 4 | 0 | 675.86904600 | 675.86879520 | 0.00025080 | 0.0033 |
| 5 | 2 | 4 | 2 | 675.86904600 | 675.86948094 | -0.00043494 | 0.0033 |
| 5 | 1 | 4 | 1 | 675.86904600 | 675.86896663 | 0.00007937 | 0.0033 |
| 5 | 3 | 4 | 3 | 675.87012900 | 675.87033813 | -0.00020913 | 0.0100 |
| 5 | 4 | 4 | 4 | 675.87141300 | 675.87153821 | -0.00012521 | 0.0100 |
| 6 | 0 | 5 | 0 | 676.24695800 | 676.24670350 | 0.00025450 | 0.0033 |
| 6 | 2 | 5 | 2 | 676.24695800 | 676.24738976 | -0.00043176 | 0.0033 |
| 6 | 1 | 5 | 1 | 676.24695800 | 676.24687506 | 0.00008294 | 0.0033 |
| 6 | 3 | 5 | 3 | 676.24807800 | 676.24824759 | -0.00016959 | 0.0100 |
| 6 | 4 | 5 | 4 | 676.24943000 | 676.24944857 | -0.00001857 | 0.0100 |
| 6 | 5 | 5 | 5 | 676.25094400 | 676.25099272 | -0.00004872 | 0.0100 |
| 7 | 0 | 6 | 0 | 676.62457400 | 676.62433591 | 0.00023809 | 0.0033 |
| 7 | 1 | 6 | 1 | 676.62457400 | 676.62450760 | 0.00006640 | 0.0033 |
| 7 | 2 | 6 | 2 | 676.62457400 | 676.62502268 | -0.00044868 | 0.0033 |
| 7 | 3 | 6 | 3 | 676.62567700 | 676.62588115 | -0.00020415 | 0.0100 |
| 7 | 4 | 6 | 4 | 676.62702000 | 676.62708303 | -0.00006303 | 0.0100 |
| 7 | 5 | 6 | 5 | 676.62870788 | 676.62862833 | 0.00007955 | 1.0000 |
| 7 | 6 | 6 | 6 | 676.63046360 | 676.63051708 | -0.00005348 | 1.0000 |
| 8 | 0 | 7 | 0 | 677.00197100 | 677.00169150 | 0.00027950 | 0.0033 |
| 8 | 1 | 7 | 1 | 677.00197100 | 677.00186332 | 0.00010768 | 0.0033 |
| 8 | 2 | 7 | 2 | 677.00197100 | 677.00237878 | -0.00040778 | 0.0033 |
| 8 | 3 | 7 | 3 | 677.00310700 | 677.00323789 | -0.00013089 | 0.0100 |
| 8 | 4 | 7 | 4 | 677.00426100 | 677.00444066 | -0.00017966 | 0.0100 |
| 8 | 5 | 7 | 5 | 677.00599574 | 677.00598711 | 0.00000863 | 1.0000 |
| 8 | 6 | 7 | 6 | 677.00791100 | 677.00787727 | 0.00003373 | 0.0100 |
| 8 | 7 | 7 | 7 | 677.01003686 | 677.01011116 | -0.00007430 | 1.0000 |
| 9 | 0 | 8 | 0 | 677.37899200 | 677.37876935 | 0.00022265 | 0.0033 |
| 9 | 1 | 8 | 1 | 677.37899200 | 677.37894129 | 0.00005071 | 0.0033 |
| 9 | 2 | 8 | 2 | 677.37899200 | 677.37945714 | -0.00046514 | 0.0033 |
| 9 | 3 | 8 | 3 | 677.38015700 | 677.38031688 | -0.00015988 | 0.0100 |
| 9 | 4 | 8 | 4 | 677.38147600 | 677.38152054 | -0.00004454 | 0.0100 |
| 9 | 5 | 8 | 5 | 677.38308347 | 677.38306814 | 0.00001533 | 1.0000 |
| 9 | 6 | 8 | 6 | 677.38495406 | 677.38495970 | -0.00000564 | 1.0000 |
| 9 | 7 | 8 | 7 | 677.38717956 | 677.38719523 | -0.00001567 | 1.0000 |
| 9 | 8 | 8 | 8 | 677.38975483 | 677.38977479 | -0.00001996 | 1.0000 |
| 10 | 0 | 9 | 0 | 677.75584400 | 677.75556850 | 0.00027550 | 0.0033 |
| 10 | 2 | 9 | 2 | 677.75584400 | 677.75625680 | -0.00041280 | 0.0033 |
| 10 | 1 | 9 | 1 | 677.75584400 | 677.75574058 | 0.00010342 | 0.0033 |
| 10 | 3 | 9 | 3 | 677.75703300 | 677.75711718 | -0.00008418 | 0.0100 |
| 10 | 4 | 9 | 4 | 677.75812100 | 677.75832173 | -0.00020073 | 0.0100 |
| 10 | 5 | 9 | 5 | 677.75987231 | 677.75987047 | 0.00000184 | 1.0000 |
| 10 | 6 | 9 | 6 | 677.76177450 | 677.76176341 | 0.00001109 | 1.0000 |
| 10 | 7 | 9 | 7 | 677.76396400 | 677.76400060 | -0.00003660 | 0.0100 |
| 10 | 8 | 9 | 8 | 677.76659887 | 677.76658206 | 0.00001681 | 1.0000 |
| 10 | 9 | 9 | 9 | 677.76950159 | 677.76950783 | -0.00000624 | 1.0000 |
| 11 | 0 | 10 | 0 | 678.13232400 | 678.13208801 | 0.00023599 | 0.0033 |
| 11 | 2 | 10 | 2 | 678.13232400 | 678.13277681 | -0.00045281 | 0.0033 |
| 11 | 1 | 10 | 1 | 678.13232400 | 678.13226021 | 0.00006379 | 0.0033 |

| 11 | 3 | 10 | 3 | 678.13340500 | 678.13363782 | -0.00023282 | 0.0100 |
| 11 | 4 | 10 | 4 | 678.13481400 | 678.13484326 | -0.00002926 | 0.0100 |
| 11 | 5 | 10 | 5 | 678.13640025 | 678.13639313 | 0.00000712 | 1.0000 |
| 11 | 6 | 10 | 6 | 678.13826997 | 678.13828747 | -0.00001750 | 1.0000 |
| 11 | 7 | 10 | 7 | 678.14051568 | 678.14052629 | -0.00001061 | 1.0000 |
| 11 | 8 | 10 | 8 | 678.14310128 | 678.14310965 | -0.00000837 | 1.0000 |
| 11 | 9 | 10 | 9 | 678.14602413 | 678.14603756 | -0.00001343 | 1.0000 |
| 11 | 10 | 10 | 10 | 678.14934367 | 678.14931008 | 0.00003359 | 1.0000 |
| 12 | 0 | 11 | 0 | 678.50862300 | 678.50832691 | 0.00029609 | 0.0033 |
| 12 | 2 | 11 | 2 | 678.50862300 | 678.50901621 | -0.00039321 | 0.0033 |
| 12 | 1 | 11 | 1 | 678.50862300 | 678.50849923 | 0.00012377 | 0.0033 |
| 12 | 3 | 11 | 3 | 678.50975100 | 678.50987785 | -0.00012685 | 0.0100 |
| 12 | 4 | 11 | 4 | 678.51105500 | 678.51108416 | -0.00002916 | 0.0100 |
| 12 | 5 | 11 | 5 | 678.51265241 | 678.51263517 | 0.00001724 | 1.0000 |
| 12 | 6 | 11 | 6 | 678.51452983 | 678.51453089 | -0.00000106 | 1.0000 |
| 12 | 7 | 11 | 7 | 678.51674900 | 678.51677135 | -0.00002235 | 0.0100 |
| 12 | 8 | 11 | 8 | 678.51932347 | 678.51935659 | -0.00003312 | 1.0000 |
| 12 | 9 | 11 | 9 | 678.52226729 | 678.52228664 | -0.00001935 | 1.0000 |
| 12 | 10 | 11 | 10 | 678.52551193 | 678.52556155 | -0.00004962 | 1.0000 |
| 12 | 11 | 11 | 11 | 678.52927988 | 678.52918136 | 0.00009852 | 1.0000 |
| 13 | 0 | 12 | 0 | 678.88451800 | 678.88428421 | 0.00023379 | 0.0033 |
| 13 | 2 | 12 | 2 | 678.88451800 | 678.88497401 | -0.00045601 | 0.0033 |
| 13 | 1 | 12 | 1 | 678.88451800 | 678.88445666 | 0.00006134 | 0.0033 |
| 13 | 3 | 12 | 3 | 678.88558400 | 678.88583628 | -0.00025228 | 0.0100 |
| 13 | 4 | 12 | 4 | 678.88699400 | 678.88704347 | -0.00004947 | 0.0100 |
| 13 | 5 | 12 | 5 | 678.88860537 | 678.88859560 | 0.00000977 | 1.0000 |
| 13 | 6 | 12 | 6 | 678.89049281 | 678.89049270 | 0.00000011 | 1.0000 |
| 13 | 7 | 12 | 7 | 678.89274651 | 678.89273479 | 0.00001172 | 1.0000 |
| 13 | 8 | 12 | 8 | 678.89529787 | 678.89532191 | -0.00002404 | 1.0000 |
| 13 | 9 | 12 | 9 | 678.89824600 | 678.89825409 | -0.00000809 | 0.0100 |
| 13 | 10 | 12 | 10 | 678.90151000 | 678.90153138 | -0.00002138 | 0.0100 |
| 13 | 11 | 12 | 11 | 678.90513200 | 678.90515382 | -0.00002182 | 0.0100 |
| 13 | 12 | 12 | 12 | 678.91000700 | 678.90912148 | 0.00088552 | 0.0050 |
| 14 | 0 | 13 | 0 | 679.26030500 | 679.25995893 | 0.00034607 | 0.0033 |
| 14 | 2 | 13 | 2 | 679.26030500 | 679.26064923 | -0.00034423 | 0.0033 |
| 14 | 1 | 13 | 1 | 679.26030500 | 679.26013151 | 0.00017349 | 0.0033 |
| 14 | 3 | 13 | 3 | 679.26140700 | 679.26151212 | -0.00010512 | 0.0100 |
| 14 | 4 | 13 | 4 | 679.26257100 | 679.26272019 | -0.00014919 | 0.0100 |
| 14 | 5 | 13 | 5 | 679.26428299 | 679.26427344 | 0.00000955 | 1.0000 |
| 14 | 6 | 13 | 6 | 679.26616363 | 679.26617192 | -0.00000829 | 1.0000 |
| 14 | 7 | 13 | 7 | 679.26840700 | 679.26841563 | -0.00000863 | 1.0000 |
| 14 | 8 | 13 | 8 | 679.27100571 | 679.27100462 | 0.00000109 | 1.0000 |
| 14 | 9 | 13 | 9 | 679.27393347 | 679.27393892 | -0.00000545 | 1.0000 |
| 14 | 10 | 13 | 10 | 679.27720541 | 679.27721858 | -0.00001317 | 1.0000 |
| 14 | 11 | 13 | 11 | 679.28083430 | 679.28084365 | -0.00000935 | 1.0000 |
| 14 | 12 | 13 | 12 | 679.28489600 | 679.28481418 | 0.00008182 | 0.0100 |
| 14 | 13 | 13 | 13 | 679.28925900 | 679.28913022 | 0.00012878 | 0.0100 |
| 15 | 0 | 14 | 0 | 679.63559600 | 679.63535009 | 0.00024591 | 0.0033 |
| 15 | 2 | 14 | 2 | 679.63559600 | 679.63604089 | -0.00044489 | 0.0033 |
| 15 | 1 | 14 | 1 | 679.63559600 | 679.63552279 | 0.00007321 | 0.0033 |
| 15 | 3 | 14 | 3 | 679.63680700 | 679.63690440 | -0.00009740 | 0.0100 |
| 15 | 4 | 14 | 4 | 679.63797000 | 679.63811334 | -0.00014334 | 0.0100 |
| 15 | 5 | 14 | 5 | 679.63967371 | 679.63966771 | 0.00000600 | 1.0000 |
| 15 | 6 | 14 | 6 | 679.64157289 | 679.64156755 | 0.00000534 | 1.0000 |
| 15 | 7 | 14 | 7 | 679.64380547 | 679.64381288 | -0.00000741 | 1.0000 |
| 15 | 8 | 14 | 8 | 679.64639884 | 679.64640373 | -0.00000489 | 1.0000 |
| 15 | 9 | 14 | 9 | 679.64932588 | 679.64934015 | -0.00001427 | 1.0000 |
| 15 | 10 | 14 | 10 | 679.65260752 | 679.65262218 | -0.00001466 | 1.0000 |
| 15 | 11 | 14 | 11 | 679.65623920 | 679.65624985 | -0.00001065 | 1.0000 |
| 15 | 12 | 14 | 12 | 679.65976600 | 679.66022324 | -0.00045724 | 0.0100 |
| 15 | 13 | 14 | 13 | 679.66427800 | 679.66454239 | -0.00026439 | 0.0100 |
| 15 | 14 | 14 | 14 | 679.66901700 | 679.66920738 | -0.00019038 | 0.0100 |
| 16 | 0 | 15 | 0 | 680.01078500 | 680.01045669 | 0.00032831 | 0.0033 |
| 16 | 1 | 15 | 1 | 680.01078500 | 680.01062951 | 0.00015549 | 0.0033 |
| 16 | 2 | 15 | 2 | 680.01078500 | 680.01114798 | -0.00036298 | 0.0033 |
| 16 | 3 | 15 | 3 | 680.01190000 | 680.01201211 | -0.00011211 | 0.0100 |
| 16 | 4 | 15 | 4 | 680.01307900 | 680.01322192 | -0.00014292 | 0.0100 |
| 16 | 5 | 15 | 5 | 680.01479009 | 680.01477741 | 0.00001268 | 1.0000 |
| 16 | 6 | 15 | 6 | 680.01669515 | 680.01667861 | 0.00001654 | 1.0000 |
| 16 | 7 | 15 | 7 | 680.01892957 | 680.01892555 | 0.00000402 | 1.0000 |
| 16 | 8 | 15 | 8 | 680.02149780 | 680.02151826 | -0.00002046 | 1.0000 |
| 16 | 9 | 15 | 9 | 680.02445884 | 680.02445678 | 0.00000206 | 1.0000 |
| 16 | 10 | 15 | 10 | 680.02774118 | 680.02774116 | 0.00000002 | 1.0000 |
| 16 | 11 | 15 | 11 | 680.03133613 | 680.03137144 | -0.00003531 | 1.0000 |
| 16 | 12 | 15 | 12 | 680.03536700 | 680.03534768 | 0.00001932 | 0.0100 |
| 16 | 13 | 15 | 13 | 680.03972848 | 680.03966993 | 0.00005855 | 1.0000 |
| 16 | 14 | 15 | 14 | 680.04446039 | 680.04433826 | 0.00012213 | 1.0000 |
| 16 | 15 | 15 | 15 | 680.04920015 | 680.04935273 | -0.00015258 | 1.0000 |
| 17 | 0 | 16 | 0 | 680.38557700 | 680.38527773 | 0.00029927 | 0.0033 |
| 17 | 1 | 16 | 1 | 680.38557700 | 680.38545068 | 0.00012632 | 0.0033 |
| 17 | 2 | 16 | 2 | 680.38557700 | 680.38596952 | -0.00039252 | 0.0033 |
| 17 | 3 | 16 | 3 | 680.38661700 | 680.38683427 | -0.00021727 | 0.0100 |
| 17 | 4 | 16 | 4 | 680.38797700 | 680.38804493 | -0.00006793 | 0.0100 |
| 17 | 5 | 16 | 5 | 680.38960629 | 680.38960153 | 0.00000476 | 1.0000 |
| 17 | 6 | 16 | 6 | 680.39149895 | 680.39150409 | -0.00000514 | 1.0000 |
| 17 | 7 | 16 | 7 | 680.39374752 | 680.39375264 | -0.00000512 | 1.0000 |
| 17 | 8 | 16 | 8 | 680.39635941 | 680.39634720 | 0.00001221 | 1.0000 |
| 17 | 9 | 16 | 9 | 680.39928742 | 680.39928782 | -0.00000040 | 1.0000 |
| 17 | 10 | 16 | 10 | 680.40257273 | 680.40257455 | -0.00000182 | 1.0000 |
| 17 | 11 | 16 | 11 | 680.40617662 | 680.40620742 | -0.00003080 | 1.0000 |
| 17 | 12 | 16 | 12 | 680.41019317 | 680.41018649 | 0.00000668 | 1.0000 |
| 17 | 13 | 16 | 13 | 680.41452322 | 680.41451183 | 0.00001139 | 1.0000 |
| 17 | 14 | 16 | 14 | 680.41922800 | 680.41918349 | 0.00004451 | 0.0100 |
| 17 | 15 | 16 | 15 | 680.42423515 | 680.42420154 | 0.00003361 | 1.0000 |
| 17 | 16 | 16 | 16 | 680.42981802 | 680.42956606 | 0.00025196 | 1.0000 |
| 18 | 0 | 17 | 0 | 680.76015100 | 680.75981222 | 0.00033878 | 0.0033 |
| 18 | 1 | 17 | 1 | 680.76015100 | 680.75998529 | 0.00016571 | 0.0033 |
| 18 | 2 | 17 | 2 | 680.76015100 | 680.76050450 | -0.00035350 | 0.0033 |
| 18 | 3 | 17 | 3 | 680.76127800 | 680.76136987 | -0.00009187 | 0.0100 |
| 18 | 4 | 17 | 4 | 680.76257100 | 680.76258139 | -0.00001039 | 0.0100 |
| 18 | 5 | 17 | 5 | 680.76416288 | 680.76413910 | 0.00002378 | 1.0000 |
| 18 | 6 | 17 | 6 | 680.76604703 | 680.76604301 | 0.00000402 | 1.0000 |
| 18 | 7 | 17 | 7 | 680.76828530 | 680.76829315 | -0.00000785 | 1.0000 |
| 18 | 8 | 17 | 8 | 680.77089814 | 680.77088956 | 0.00000858 | 1.0000 |
| 18 | 9 | 17 | 9 | 680.77383052 | 680.77383227 | -0.00000175 | 1.0000 |
| 18 | 10 | 17 | 10 | 680.77712629 | 680.77712133 | 0.00000496 | 1.0000 |
| 18 | 11 | 17 | 11 | 680.78098300 | 680.78075679 | 0.00022621 | 0.0100 |
| 18 | 12 | 17 | 12 | 680.78474486 | 680.78473869 | 0.00000617 | 1.0000 |
| 18 | 13 | 17 | 13 | 680.78900540 | 680.78906710 | -0.00006170 | 1.0000 |
| 18 | 14 | 17 | 14 | 680.79370661 | 680.79374208 | -0.00003547 | 1.0000 |
| 18 | 15 | 17 | 15 | 680.79874459 | 680.79876370 | -0.00001911 | 1.0000 |
| 18 | 16 | 17 | 16 | 680.80396663 | 680.80413203 | -0.00016540 | 1.0000 |
| 18 | 17 | 17 | 17 | 680.81001348 | 680.80984715 | 0.00016633 | 1.0000 |
| 19 | 0 | 18 | 0 | 681.13434200 | 681.13405917 | 0.00028283 | 0.0033 |
| 19 | 2 | 18 | 2 | 681.13434200 | 681.13475194 | -0.00040994 | 0.0033 |
| 19 | 1 | 18 | 1 | 681.13434200 | 681.13423236 | 0.00010964 | 0.0033 |
| 19 | 3 | 18 | 3 | 681.13543700 | 681.13561792 | -0.00018092 | 0.0100 |
| 19 | 4 | 18 | 4 | 681.13675900 | 681.13683030 | -0.00007130 | 0.0100 |
| 19 | 5 | 18 | 5 | 681.13839905 | 681.13838911 | 0.00000994 | 1.0000 |
| 19 | 6 | 18 | 6 | 681.14028400 | 681.14029437 | -0.00001037 | 0.0100 |
| 19 | 7 | 18 | 7 | 681.14253884 | 681.14254611 | -0.00000727 | 1.0000 |
| 19 | 8 | 18 | 8 | 681.14512170 | 681.14514435 | -0.00002265 | 1.0000 |
| 19 | 9 | 18 | 9 | 681.14810284 | 681.14808914 | 0.00001370 | 1.0000 |
| 19 | 10 | 18 | 10 | 681.15137798 | 681.15138053 | -0.00000255 | 1.0000 |
| 19 | 11 | 18 | 11 | 681.15494733 | 681.15501855 | -0.00007122 | 1.0000 |
| 19 | 12 | 18 | 12 | 681.15907625 | 681.15900327 | 0.00007298 | 1.0000 |
| 19 | 13 | 18 | 13 | 681.16336997 | 681.16333474 | 0.00003523 | 1.0000 |
| 19 | 14 | 18 | 14 | 681.16800786 | 681.16801303 | -0.00000517 | 1.0000 |
| 19 | 15 | 18 | 15 | 681.17310329 | 681.17303820 | 0.00006509 | 1.0000 |
| 19 | 16 | 18 | 16 | 681.17839418 | 681.17841033 | -0.00001615 | 1.0000 |
| 19 | 17 | 18 | 17 | 681.18403614 | 681.18412950 | -0.00009336 | 1.0000 |
| 19 | 18 | 18 | 18 | 681.19023828 | 681.19019579 | 0.00004249 | 1.0000 |
| 20 | 0 | 19 | 0 | 681.50836800 | 681.50801759 | 0.00035041 | 0.0033 |
| 20 | 2 | 19 | 2 | 681.50836800 | 681.50871085 | -0.00034285 | 0.0033 |
| 20 | 1 | 19 | 1 | 681.50836800 | 681.50819091 | 0.00017709 | 0.0033 |
| 20 | 3 | 19 | 3 | 681.50954200 | 681.50957744 | -0.00003544 | 0.0100 |
| 20 | 4 | 19 | 4 | 681.51070200 | 681.51079067 | -0.00008867 | 0.0100 |
| 20 | 5 | 19 | 5 | 681.51239692 | 681.51235058 | 0.00004634 | 1.0000 |
| 20 | 6 | 19 | 6 | 681.51427623 | 681.51425718 | 0.00001905 | 1.0000 |
| 20 | 7 | 19 | 7 | 681.51652533 | 681.51651050 | 0.00001483 | 1.0000 |
| 20 | 8 | 19 | 8 | 681.51909555 | 681.51911058 | -0.00001503 | 1.0000 |
| 20 | 9 | 19 | 9 | 681.52204404 | 681.52205745 | -0.00001341 | 1.0000 |
| 20 | 10 | 19 | 10 | 681.52534503 | 681.52535115 | -0.00000612 | 1.0000 |
| 20 | 11 | 19 | 11 | 681.52893300 | 681.52899174 | -0.00005874 | 0.0050 |
| 20 | 12 | 19 | 12 | 681.53295413 | 681.53297926 | -0.00002513 | 1.0000 |
| 20 | 13 | 19 | 13 | 681.53730080 | 681.53731378 | -0.00001298 | 1.0000 |
| 20 | 14 | 19 | 14 | 681.54200654 | 681.54199536 | 0.00001118 | 1.0000 |
| 20 | 15 | 19 | 15 | 681.54697987 | 681.54702407 | -0.00004420 | 1.0000 |
| 20 | 16 | 19 | 16 | 681.55249291 | 681.55239998 | 0.00009293 | 1.0000 |
| 20 | 17 | 19 | 17 | 681.55823512 | 681.55812318 | 0.00011194 | 1.0000 |
| 20 | 18 | 19 | 18 | 681.56375842 | 681.56419374 | -0.00043532 | 1.0000 |
| 20 | 19 | 19 | 19 | 681.57073200 | 681.57061175 | 0.00012025 | 0.0100 |
| 21 | 0 | 20 | 0 | 681.88188600 | 681.88168650 | 0.00019950 | 0.0033 |
| 21 | 2 | 20 | 2 | 681.88188600 | 681.88238025 | -0.00049425 | 0.0033 |
| 21 | 1 | 20 | 1 | 681.88188600 | 681.88185994 | 0.00002606 | 0.0033 |
| 21 | 3 | 20 | 3 | 681.88310900 | 681.88324744 | -0.00013844 | 0.0100 |
| 21 | 4 | 20 | 4 | 681.88437700 | 681.88446153 | -0.00008453 | 0.0100 |
| 21 | 5 | 20 | 5 | 681.88607339 | 681.88602253 | 0.00005086 | 1.0000 |
| 21 | 6 | 20 | 6 | 681.88795351 | 681.88793047 | 0.00002304 | 1.0000 |
| 21 | 7 | 20 | 7 | 681.89018781 | 681.89018537 | 0.00000244 | 1.0000 |
| 21 | 8 | 20 | 8 | 681.89279003 | 681.89278727 | 0.00000276 | 1.0000 |
| 21 | 9 | 20 | 9 | 681.89574156 | 681.89573620 | 0.00000536 | 1.0000 |
| 21 | 10 | 20 | 10 | 681.89903791 | 681.89903221 | 0.00000570 | 1.0000 |
| 21 | 11 | 20 | 11 | 681.90253000 | 681.90267535 | -0.00014535 | 0.0100 |
| 21 | 12 | 20 | 12 | 681.90669123 | 681.90666567 | 0.00002556 | 1.0000 |
| 21 | 13 | 20 | 13 | 681.91098391 | 681.91100323 | -0.00001932 | 1.0000 |
| 21 | 14 | 20 | 14 | 681.91579495 | 681.91568809 | 0.00010686 | 1.0000 |
| 21 | 15 | 20 | 15 | 681.92067553 | 681.92072033 | -0.00004480 | 1.0000 |
| 21 | 16 | 20 | 16 | 681.92601896 | 681.92610001 | -0.00008105 | 1.0000 |
| 21 | 17 | 20 | 17 | 681.93179117 | 681.93182721 | -0.00003604 | 1.0000 |
| 21 | 18 | 20 | 18 | 681.93791084 | 681.93790202 | 0.00000882 | 1.0000 |
| 21 | 19 | 20 | 19 | 681.94438026 | 681.94432453 | 0.00005573 | 1.0000 |
| 21 | 20 | 20 | 20 | 681.95107400 | 681.95109484 | -0.00002084 | 0.0050 |
| 22 | 0 | 21 | 0 | 682.25544200 | 682.25506493 | 0.00037707 | 0.0033 |
| 22 | 1 | 21 | 1 | 682.25544200 | 682.25523849 | 0.00020351 | 0.0033 |
| 22 | 2 | 21 | 2 | 682.25544200 | 682.25575916 | -0.00031716 | 0.0033 |
| 22 | 3 | 21 | 3 | 682.25647600 | 682.25662696 | -0.00015096 | 0.0050 |
| 22 | 4 | 21 | 4 | 682.25784100 | 682.25784190 | -0.00000090 | 0.0100 |
| 22 | 5 | 21 | 5 | 682.25943765 | 682.25940399 | 0.00003366 | 1.0000 |
| 22 | 6 | 21 | 6 | 682.26133265 | 682.26131326 | 0.00001939 | 1.0000 |
| 22 | 7 | 21 | 7 | 682.26359938 | 682.26356973 | 0.00002965 | 1.0000 |
| 22 | 8 | 21 | 8 | 682.26616458 | 682.26617345 | -0.00000887 | 1.0000 |
| 22 | 9 | 21 | 9 | 682.26912962 | 682.26912444 | 0.00000518 | 1.0000 |
| 22 | 10 | 21 | 10 | 682.27248200 | 682.27242275 | 0.00005925 | 0.0100 |
| 22 | 11 | 21 | 11 | 682.27613390 | 682.27606843 | 0.00006547 | 1.0000 |
| 22 | 12 | 21 | 12 | 682.28021600 | 682.28006153 | 0.00015447 | 0.0100 |

| 22 | 13 | 21 | 13 | 682.28443414 | 682.28440212 | 0.00003202 | 1.0000 |
|---|---|---|---|---|---|---|---|
| 22 | 14 | 21 | 14 | 682.28905663 | 682.28909025 | -0.00003362 | 1.0000 |
| 22 | 15 | 21 | 15 | 682.29416584 | 682.29412600 | 0.00003984 | 1.0000 |
| 22 | 16 | 21 | 16 | 682.29957031 | 682.29950943 | 0.00006088 | 1.0000 |
| 22 | 17 | 21 | 17 | 682.30528747 | 682.30524063 | 0.00004684 | 1.0000 |
| 22 | 18 | 21 | 18 | 682.31140752 | 682.31131968 | 0.00008784 | 1.0000 |
| 22 | 19 | 21 | 19 | 682.31762569 | 682.31774667 | -0.00012098 | 1.0000 |
| 22 | 20 | 21 | 20 | 682.32468219 | 682.32452169 | 0.00016050 | 1.0000 |
| 22 | 21 | 21 | 21 | 682.33139871 | 682.33164485 | -0.00024614 | 1.0000 |
| 23 | 0 | 22 | 0 | 682.62844200 | 682.62815193 | 0.00029007 | 0.0033 |
| 23 | 2 | 22 | 2 | 682.62844200 | 682.62884664 | -0.00040464 | 0.0033 |
| 23 | 1 | 22 | 1 | 682.62844200 | 682.62832560 | 0.00011640 | 0.0033 |
| 23 | 3 | 22 | 3 | 682.62957100 | 682.62971504 | -0.00014404 | 0.0100 |
| 23 | 4 | 22 | 4 | 682.63092600 | 682.63093082 | -0.00000482 | 0.0100 |
| 23 | 5 | 22 | 5 | 682.63249998 | 682.63249400 | 0.00000598 | 1.0000 |
| 23 | 6 | 22 | 6 | 682.63440495 | 682.63440459 | 0.00000036 | 1.0000 |
| 23 | 7 | 22 | 7 | 682.63665849 | 682.63666263 | -0.00000414 | 1.0000 |
| 23 | 8 | 22 | 8 | 682.63926055 | 682.63926816 | -0.00000761 | 1.0000 |
| 23 | 10 | 22 | 10 | 682.64561771 | 682.64552180 | 0.00009591 | 1.0000 |
| 23 | 11 | 22 | 11 | 682.64919153 | 682.64917002 | 0.00002151 | 1.0000 |
| 23 | 12 | 22 | 12 | 682.65307320 | 682.65316589 | -0.00009269 | 1.0000 |
| 23 | 13 | 22 | 13 | 682.65752704 | 682.65750949 | 0.00001755 | 1.0000 |
| 23 | 14 | 22 | 14 | 682.66222210 | 682.66220088 | 0.00002122 | 1.0000 |
| 23 | 15 | 22 | 15 | 682.66721275 | 682.66724012 | -0.00002737 | 1.0000 |
| 23 | 16 | 22 | 16 | 682.67253200 | 682.67262729 | -0.00009529 | 0.0100 |
| 23 | 17 | 22 | 17 | 682.67837215 | 682.67836247 | 0.00000968 | 1.0000 |
| 23 | 18 | 22 | 18 | 682.68446923 | 682.68444574 | 0.00002349 | 1.0000 |
| 23 | 19 | 22 | 19 | 682.69104700 | 682.69087719 | 0.00016981 | 0.0100 |
| 23 | 20 | 22 | 20 | 682.69760017 | 682.69765692 | -0.00005675 | 1.0000 |
| 23 | 21 | 22 | 21 | 682.70484037 | 682.70478503 | 0.00005534 | 1.0000 |
| 23 | 22 | 22 | 22 | 682.71235300 | 682.71226161 | 0.00009139 | 0.0100 |

| 24 | 0  | 23 | 0  | 683.00135200 | 683.00094654 | 0.00040546  | 0.0033 |
|----|----|----|----|--------------|--------------|-------------|--------|
| 24 | 1  | 23 | 1  | 683.00135200 | 683.00112034 | 0.00023166  | 0.0033 |
| 24 | 2  | 23 | 2  | 683.00135200 | 683.00164174 | -0.00028974 | 0.0033 |
| 24 | 3  | 23 | 3  | 683.00247500 | 683.00251074 | -0.00003574 | 0.0100 |
| 24 | 4  | 23 | 4  | 683.00371000 | 683.00372736 | -0.00001736 | 0.0100 |
| 24 | 5  | 23 | 5  | 683.00531188 | 683.00529162 | 0.00002026  | 1.0000 |
| 24 | 6  | 23 | 6  | 683.00719380 | 683.00720353 | -0.00000973 | 1.0000 |
| 24 | 7  | 23 | 7  | 683.00947277 | 683.00946314 | 0.00000963  | 1.0000 |
| 24 | 8  | 23 | 8  | 683.01206706 | 683.01207046 | -0.00000340 | 1.0000 |
| 24 | 9  | 23 | 9  | 683.01500308 | 683.01502555 | -0.00002247 | 1.0000 |
| 24 | 10 | 23 | 10 | 683.01818942 | 683.01832843 | -0.00013901 | 1.0000 |
| 24 | 11 | 23 | 11 | 683.02205123 | 683.02197917 | 0.00007206  | 1.0000 |
| 24 | 12 | 23 | 12 | 683.02606217 | 683.02597781 | 0.00008436  | 1.0000 |
| 24 | 13 | 23 | 13 | 683.03030803 | 683.03032441 | -0.00001638 | 1.0000 |
| 24 | 14 | 23 | 14 | 683.03503423 | 683.03501904 | 0.00001519  | 1.0000 |
| 24 | 15 | 23 | 15 | 683.04009338 | 683.04006177 | 0.00003161  | 1.0000 |
| 24 | 16 | 23 | 16 | 683.04544511 | 683.04545266 | -0.00000755 | 1.0000 |
| 24 | 17 | 23 | 17 | 683.05121836 | 683.05119180 | 0.00002656  | 1.0000 |
| 24 | 18 | 23 | 18 | 683.05731608 | 683.05727928 | 0.00003680  | 1.0000 |
| 24 | 19 | 23 | 19 | 683.06374111 | 683.06371517 | 0.00002594  | 1.0000 |
| 24 | 20 | 23 | 20 | 683.07060100 | 683.07049959 | 0.00010141  | 0.0100 |
| 24 | 21 | 23 | 21 | 683.07765430 | 683.07763262 | 0.00002168  | 1.0000 |
| 24 | 22 | 23 | 22 | 683.08523321 | 683.08511436 | 0.00011885  | 1.0000 |
| 24 | 23 | 23 | 23 | 683.09298327 | 683.09294494 | 0.00003833  | 1.0000 |
| 25 | 0  | 24 | 0  | 683.37364600 | 683.37344786 | 0.00019814  | 0.0033 |
| 25 | 2  | 24 | 2  | 683.37364600 | 683.37414353 | -0.00049753 | 0.0033 |
| 25 | 1  | 24 | 1  | 683.37364600 | 683.37362178 | 0.00002422  | 0.0033 |
| 25 | 3  | 24 | 3  | 683.37496400 | 683.37501313 | -0.00004913 | 0.0100 |
| 25 | 4  | 24 | 4  | 683.37616300 | 683.37623059 | -0.00006759 | 0.0100 |
| 25 | 5  | 24 | 5  | 683.37780764 | 683.37779592 | 0.00001172  | 1.0000 |
| 25 | 6  | 24 | 6  | 683.37970959 | 683.37970916 | 0.00000043  | 1.0000 |
| 25 | 7 | 24 | 7 | 683.38196299 | 683.38197032 | -0.00000733 | 1.0000 |
| 25 | 8 | 24 | 8 | 683.38457566 | 683.38457944 | -0.00000378 | 1.0000 |
| 25 | 9 | 24 | 9 | 683.38752977 | 683.38753655 | -0.00000678 | 1.0000 |
| 25 | 10 | 24 | 10 | 683.39090800 | 683.39084171 | 0.00006629 | 0.0100 |
| 25 | 11 | 24 | 11 | 683.39449765 | 683.39449496 | 0.00000269 | 1.0000 |
| 25 | 12 | 24 | 12 | 683.39849753 | 683.39849635 | 0.00000118 | 1.0000 |
| 25 | 13 | 24 | 13 | 683.40285846 | 683.40284595 | 0.00001251 | 1.0000 |
| 25 | 14 | 24 | 14 | 683.40752791 | 683.40754381 | -0.00001590 | 1.0000 |
| 25 | 15 | 24 | 15 | 683.41257082 | 683.41259000 | -0.00001918 | 1.0000 |
| 25 | 16 | 24 | 16 | 683.41798711 | 683.41798461 | 0.00000250 | 1.0000 |
| 25 | 17 | 24 | 17 | 683.42369478 | 683.42372770 | -0.00003292 | 1.0000 |
| 25 | 18 | 24 | 18 | 683.42988073 | 683.42981936 | 0.00006137 | 1.0000 |
| 25 | 19 | 24 | 19 | 683.43622350 | 683.43625969 | -0.00003619 | 1.0000 |
| 25 | 20 | 24 | 20 | 683.44304319 | 683.44304876 | -0.00000557 | 1.0000 |
| 25 | 21 | 24 | 21 | 683.45016853 | 683.45018670 | -0.00001817 | 1.0000 |
| 25 | 22 | 24 | 22 | 683.45775731 | 683.45767359 | 0.00008372 | 1.0000 |
| 25 | 23 | 24 | 23 | 683.46547324 | 683.46550955 | -0.00003631 | 1.0000 |
| 25 | 24 | 24 | 24 | 683.47380800 | 683.47369470 | 0.00011330 | 0.0100 |
| 26 | 0 | 25 | 0 | 683.74609400 | 683.74565498 | 0.00043902 | 0.0033 |
| 26 | 2 | 25 | 2 | 683.74609400 | 683.74635113 | -0.00025713 | 0.0033 |
| 26 | 1 | 25 | 1 | 683.74609400 | 683.74582901 | 0.00026499 | 0.0033 |
| 26 | 3 | 25 | 3 | 683.74705600 | 683.74722132 | -0.00016532 | 0.0100 |
| 26 | 4 | 25 | 4 | 683.74841300 | 683.74843961 | -0.00002661 | 0.0100 |
| 26 | 5 | 25 | 5 | 683.75001819 | 683.75000602 | 0.00001217 | 1.0000 |
| 26 | 6 | 25 | 6 | 683.75191750 | 683.75192056 | -0.00000306 | 1.0000 |
| 26 | 7 | 25 | 7 | 683.75416577 | 683.75418327 | -0.00001750 | 1.0000 |
| 26 | 8 | 25 | 8 | 683.75679197 | 683.75679418 | -0.00000221 | 1.0000 |
| 26 | 9 | 25 | 9 | 683.75973952 | 683.75975332 | -0.00001380 | 1.0000 |
| 26 | 10 | 25 | 10 | 683.76304951 | 683.76306075 | -0.00001124 | 1.0000 |
| 26 | 11 | 25 | 11 | 683.76668485 | 683.76671650 | -0.00003165 | 1.0000 |
| 26 | 12 | 25 | 12 | 683.77070222 | 683.77072063 | -0.00001841 | 1.0000 |

| 26 | 13 | 25 | 13 | 683.77510961 | 683.77507321 | 0.00003640 | 1.0000 |
|----|----|----|----|--------------|--------------|-------------|--------|
| 26 | 14 | 25 | 14 | 683.77977672 | 683.77977429 | 0.00000243 | 1.0000 |
| 26 | 15 | 25 | 15 | 683.78483947 | 683.78482394 | 0.00001553 | 1.0000 |
| 26 | 16 | 25 | 16 | 683.79023298 | 683.79022224 | 0.00001074 | 1.0000 |
| 26 | 17 | 25 | 17 | 683.79606170 | 683.79596926 | 0.00009244 | 1.0000 |
| 26 | 18 | 25 | 18 | 683.80202731 | 683.80206509 | -0.00003778 | 1.0000 |
| 26 | 19 | 25 | 19 | 683.80851600 | 683.80850983 | 0.00000617 | 1.0000 |
| 26 | 20 | 25 | 20 | 683.81537300 | 683.81530355 | 0.00006945 | 0.0100 |
| 26 | 21 | 25 | 21 | 683.82249035 | 683.82244637 | 0.00004398 | 1.0000 |
| 26 | 22 | 25 | 22 | 683.82988927 | 683.82993839 | -0.00004912 | 1.0000 |
| 26 | 23 | 25 | 23 | 683.83787300 | 683.83777972 | 0.00009328 | 0.0100 |
| 26 | 24 | 25 | 24 | 683.84612298 | 683.84597046 | 0.00015252 | 1.0000 |
| 26 | 25 | 25 | 25 | 683.85425121 | 683.85451075 | -0.00025954 | 1.0000 |
| 27 | 0  | 26 | 0  | 684.11774700 | 684.11756702 | 0.00017998 | 0.1667 |
| 27 | 2  | 26 | 2  | 684.11774700 | 684.11826364 | -0.00051664 | 0.0017 |
| 27 | 1  | 26 | 1  | 684.11774700 | 684.11774118 | 0.00000582 | 0.0017 |
| 27 | 3  | 26 | 3  | 684.11905400 | 684.11913443 | -0.00008043 | 0.0100 |
| 27 | 4  | 26 | 4  | 684.12022800 | 684.12035356 | -0.00012556 | 0.0100 |
| 27 | 5  | 26 | 5  | 684.12191637 | 684.12192103 | -0.00000466 | 1.0000 |
| 27 | 6  | 26 | 6  | 684.12381400 | 684.12383688 | -0.00002288 | 1.0000 |
| 27 | 7  | 26 | 7  | 684.12607929 | 684.12610113 | -0.00002184 | 1.0000 |
| 27 | 8  | 26 | 8  | 684.12869908 | 684.12871382 | -0.00001474 | 1.0000 |
| 27 | 9  | 26 | 9  | 684.13153165 | 684.13167498 | -0.00014333 | 1.0000 |
| 27 | 10 | 26 | 10 | 684.13502876 | 684.13498466 | 0.00004410 | 1.0000 |
| 27 | 11 | 26 | 11 | 684.13867412 | 684.13864291 | 0.00003121 | 1.0000 |
| 27 | 12 | 26 | 12 | 684.14266272 | 684.14264977 | 0.00001295 | 1.0000 |
| 27 | 13 | 26 | 13 | 684.14698268 | 684.14700531 | -0.00002263 | 1.0000 |
| 27 | 14 | 26 | 14 | 684.15166360 | 684.15170960 | -0.00004600 | 1.0000 |
| 27 | 15 | 26 | 15 | 684.15677738 | 684.15676269 | 0.00001469 | 1.0000 |
| 27 | 16 | 26 | 16 | 684.16215600 | 684.16216467 | -0.00000867 | 0.0100 |
| 27 | 17 | 26 | 17 | 684.16789080 | 684.16791561 | -0.00002481 | 1.0000 |
| 27 | 18 | 26 | 18 | 684.17402092 | 684.17401560 | 0.00000532 | 1.0000 |
| 27 | 19 | 26 | 19 | 684.18043850 | 684.18046473 | -0.00002623 | 1.0000 |
| 27 | 20 | 26 | 20 | 684.18727494 | 684.18726308 | 0.00001186 | 1.0000 |
| 27 | 21 | 26 | 21 | 684.19435590 | 684.19441077 | -0.00005487 | 1.0000 |
| 27 | 22 | 26 | 22 | 684.20186552 | 684.20190789 | -0.00004237 | 1.0000 |
| 27 | 23 | 26 | 23 | 684.20981825 | 684.20975456 | 0.00006369 | 1.0000 |
| 27 | 24 | 26 | 24 | 684.21780671 | 684.21795088 | -0.00014417 | 1.0000 |
| 27 | 25 | 26 | 25 | 684.22648900 | 684.22649699 | -0.00000799 | 0.0100 |
| 27 | 26 | 26 | 26 | 684.23542400 | 684.23539300 | 0.00003100 | 0.0100 |
| 28 | 0 | 27 | 0 | 684.48948100 | 684.48918314 | 0.00029786 | 0.0033 |
| 28 | 2 | 27 | 2 | 684.48948100 | 684.48988024 | -0.00039924 | 0.0033 |
| 28 | 1 | 27 | 1 | 684.48948100 | 684.48935742 | 0.00012358 | 0.0033 |
| 28 | 3 | 27 | 3 | 684.49062300 | 684.49075162 | -0.00012862 | 0.0100 |
| 28 | 4 | 27 | 4 | 684.49189600 | 684.49197157 | -0.00007557 | 0.0100 |
| 28 | 5 | 27 | 5 | 684.49353445 | 684.49354011 | -0.00000566 | 1.0000 |
| 28 | 6 | 27 | 6 | 684.49544265 | 684.49545726 | -0.00001461 | 1.0000 |
| 28 | 7 | 27 | 7 | 684.49776600 | 684.49772305 | 0.00004295 | 0.0100 |
| 28 | 8 | 27 | 8 | 684.50033101 | 684.50033751 | -0.00000650 | 1.0000 |
| 28 | 9 | 27 | 9 | 684.50327558 | 684.50330069 | -0.00002511 | 1.0000 |
| 28 | 10 | 27 | 10 | 684.50657403 | 684.50661261 | -0.00003858 | 1.0000 |
| 28 | 11 | 27 | 11 | 684.51025271 | 684.51027334 | -0.00002063 | 1.0000 |
| 28 | 12 | 27 | 12 | 684.51426949 | 684.51428292 | -0.00001343 | 1.0000 |
| 28 | 13 | 27 | 13 | 684.51862494 | 684.51864142 | -0.00001648 | 1.0000 |
| 28 | 14 | 27 | 14 | 684.52334900 | 684.52334890 | 0.00000010 | 1.0000 |
| 28 | 15 | 27 | 15 | 684.52837633 | 684.52840542 | -0.00002909 | 1.0000 |
| 28 | 16 | 27 | 16 | 684.53384046 | 684.53381106 | 0.00002940 | 1.0000 |
| 28 | 17 | 27 | 17 | 684.53952336 | 684.53956591 | -0.00004255 | 1.0000 |
| 28 | 18 | 27 | 18 | 684.54565585 | 684.54567003 | -0.00001418 | 1.0000 |
| 28 | 19 | 27 | 19 | 684.55214965 | 684.55212353 | 0.00002612 | 1.0000 |
| 28 | 20 | 27 | 20 | 684.55890534 | 684.55892650 | -0.00002116 | 1.0000 |
| 28 | 21 | 27 | 21 | 684.56607659 | 684.56607904 | -0.00000245 | 1.0000 |

| 28 | 22 | 27 | 22 | 684.57367785 | 684.57358124 | 0.00009661 | 1.0000 |
|---|---|---|---|---|---|---|---|
| 28 | 23 | 27 | 23 | 684.58136992 | 684.58143323 | -0.00006331 | 1.0000 |
| 28 | 24 | 27 | 24 | 684.58966200 | 684.58963511 | 0.00002689 | 0.0100 |
| 28 | 25 | 27 | 25 | 684.59820197 | 684.59818701 | 0.00001496 | 1.0000 |
| 28 | 26 | 27 | 26 | 684.60705851 | 684.60708905 | -0.00003054 | 1.0000 |
| 28 | 27 | 27 | 27 | 684.61616200 | 684.61634136 | -0.00017936 | 0.0100 |
| 29 | 0 | 28 | 0 | 684.86071600 | 684.86050253 | 0.00021347 | 0.0033 |
| 29 | 2 | 28 | 2 | 684.86071600 | 684.86120010 | -0.00048410 | 0.0033 |
| 29 | 1 | 28 | 1 | 684.86071600 | 684.86067693 | 0.00003907 | 0.0033 |
| 29 | 3 | 28 | 3 | 684.86198300 | 684.86207207 | -0.00008907 | 0.0100 |
| 29 | 4 | 28 | 4 | 684.86317400 | 684.86329285 | -0.00011885 | 0.0050 |
| 29 | 5 | 28 | 5 | 684.86486847 | 684.86486245 | 0.00000602 | 1.0000 |
| 29 | 6 | 28 | 6 | 684.86676407 | 684.86678089 | -0.00001682 | 1.0000 |
| 29 | 7 | 28 | 7 | 684.86904219 | 684.86904821 | -0.00000602 | 1.0000 |
| 29 | 8 | 28 | 8 | 684.87163731 | 684.87166444 | -0.00002713 | 1.0000 |
| 29 | 9 | 28 | 9 | 684.87456500 | 684.87462962 | -0.00006462 | 0.0100 |
| 29 | 10 | 28 | 10 | 684.87796600 | 684.87794378 | 0.00002222 | 0.0100 |
| 29 | 11 | 28 | 11 | 684.88155523 | 684.88160698 | -0.00005175 | 1.0000 |
| 29 | 12 | 28 | 12 | 684.88561740 | 684.88561927 | -0.00000187 | 1.0000 |
| 29 | 13 | 28 | 13 | 684.88994692 | 684.88998071 | -0.00003379 | 1.0000 |
| 29 | 14 | 28 | 14 | 684.89470484 | 684.89469137 | 0.00001347 | 1.0000 |
| 29 | 15 | 28 | 15 | 684.89977281 | 684.89975131 | 0.00002150 | 1.0000 |
| 29 | 16 | 28 | 16 | 684.90510717 | 684.90516060 | -0.00005343 | 1.0000 |
| 29 | 17 | 28 | 17 | 684.91091240 | 684.91091933 | -0.00000693 | 1.0000 |
| 29 | 18 | 28 | 18 | 684.91702500 | 684.91702758 | -0.00000258 | 0.0100 |
| 29 | 19 | 28 | 19 | 684.92347569 | 684.92348544 | -0.00000975 | 1.0000 |
| 29 | 20 | 28 | 20 | 684.93023991 | 684.93029300 | -0.00005309 | 1.0000 |
| 29 | 21 | 28 | 21 | 684.93750415 | 684.93745037 | 0.00005378 | 1.0000 |
| 29 | 22 | 28 | 22 | 684.94478000 | 684.94495764 | -0.00017764 | 0.0100 |
| 29 | 23 | 28 | 23 | 684.95284441 | 684.95281492 | 0.00002949 | 1.0000 |
| 29 | 24 | 28 | 24 | 684.96103202 | 684.96102234 | 0.00000968 | 1.0000 |
| 29 | 25 | 28 | 25 | 684.96963612 | 684.96958001 | 0.00005611 | 1.0000 |
| 29 | 26 | 28 | 26 | 684.97855800 | 684.97848806 | 0.00006994 | 0.0100 |
| 29 | 27 | 28 | 27 | 684.98771900 | 684.98774661 | -0.00002761 | 0.0100 |
| 29 | 28 | 28 | 28 | 684.99730200 | 684.99735580 | -0.00005380 | 0.0100 |
| 30 | 0 | 29 | 0 | 685.23188600 | 685.23152442 | 0.00036158 | 0.0017 |
| 30 | 2 | 29 | 2 | 685.23188600 | 685.23222245 | -0.00033645 | 0.0017 |
| 30 | 1 | 29 | 1 | 685.23188600 | 685.23169893 | 0.00018707 | 0.0017 |
| 30 | 3 | 29 | 3 | 685.23300700 | 685.23309501 | -0.00008801 | 0.0100 |
| 30 | 4 | 29 | 4 | 685.23426000 | 685.23431661 | -0.00005661 | 0.0100 |
| 30 | 5 | 29 | 5 | 685.23590328 | 685.23588726 | 0.00001602 | 1.0000 |
| 30 | 6 | 29 | 6 | 685.23780532 | 685.23780700 | -0.00000168 | 1.0000 |
| 30 | 7 | 29 | 7 | 685.24005963 | 685.24007584 | -0.00001621 | 1.0000 |
| 30 | 8 | 29 | 8 | 685.24266352 | 685.24269383 | -0.00003031 | 1.0000 |
| 30 | 9 | 29 | 9 | 685.24570784 | 685.24566100 | 0.00004684 | 1.0000 |
| 30 | 10 | 29 | 10 | 685.24898610 | 685.24897739 | 0.00000871 | 1.0000 |
| 30 | 11 | 29 | 11 | 685.25269998 | 685.25264305 | 0.00005693 | 1.0000 |
| 30 | 12 | 29 | 12 | 685.25660182 | 685.25665805 | -0.00005623 | 1.0000 |
| 30 | 13 | 29 | 13 | 685.26104285 | 685.26102242 | 0.00002043 | 1.0000 |
| 30 | 14 | 29 | 14 | 685.26568057 | 685.26573624 | -0.00005567 | 1.0000 |
| 30 | 15 | 29 | 15 | 685.27080300 | 685.27079958 | 0.00000342 | 1.0000 |
| 30 | 16 | 29 | 16 | 685.27621764 | 685.27621251 | 0.00000513 | 1.0000 |
| 30 | 17 | 29 | 17 | 685.28195344 | 685.28197511 | -0.00002167 | 1.0000 |
| 30 | 18 | 29 | 18 | 685.28810771 | 685.28808747 | 0.00002024 | 1.0000 |
| 30 | 19 | 29 | 19 | 685.29454124 | 685.29454967 | -0.00000843 | 1.0000 |
| 30 | 20 | 29 | 20 | 685.30143754 | 685.30136180 | 0.00007574 | 1.0000 |
| 30 | 21 | 29 | 21 | 685.30851034 | 685.30852398 | -0.00001364 | 1.0000 |
| 30 | 22 | 29 | 22 | 685.31604990 | 685.31603629 | 0.00001361 | 1.0000 |
| 30 | 23 | 29 | 23 | 685.32390593 | 685.32389886 | 0.00000707 | 1.0000 |
| 30 | 24 | 29 | 24 | 685.33212665 | 685.33211179 | 0.00001486 | 1.0000 |
| 30 | 25 | 29 | 25 | 685.34068000 | 685.34067521 | 0.00000479 | 0.0100 |
| 30 | 26 | 29 | 26 | 685.34960040 | 685.34958923 | 0.00001117 | 1.0000 |
| 30 | 27 | 29 | 27 | 685.35884772 | 685.35885400 | -0.00000628 | 1.0000 |
| 30 | 28 | 29 | 28 | 685.36846100 | 685.36846965 | -0.00000865 | 0.0100 |
| 30 | 29 | 29 | 29 | 685.37841800 | 685.37843631 | -0.00001831 | 0.0100 |
| 31 | 0 | 30 | 0 | 685.60253800 | 685.60224805 | 0.00028995 | 0.0033 |
| 31 | 2 | 30 | 2 | 685.60253800 | 685.60294656 | -0.00040856 | 0.0033 |
| 31 | 1 | 30 | 1 | 685.60253800 | 685.60242268 | 0.00011532 | 0.0033 |
| 31 | 3 | 30 | 3 | 685.60355500 | 685.60381970 | -0.00026470 | 0.0100 |
| 31 | 4 | 30 | 4 | 685.60495100 | 685.60504211 | -0.00009111 | 0.0100 |
| 31 | 5 | 30 | 5 | 685.60661413 | 685.60661382 | 0.00000031 | 1.0000 |
| 31 | 6 | 30 | 6 | 685.60852918 | 685.60853484 | -0.00000566 | 1.0000 |
| 31 | 7 | 30 | 7 | 685.61080191 | 685.61080520 | -0.00000329 | 1.0000 |
| 31 | 8 | 30 | 8 | 685.61360600 | 685.61342494 | 0.00018106 | 0.0100 |
| 31 | 9 | 30 | 9 | 685.61633146 | 685.61639410 | -0.00006264 | 1.0000 |
| 31 | 10 | 30 | 10 | 685.61966100 | 685.61971271 | -0.00005171 | 1.0000 |
| 31 | 11 | 30 | 11 | 685.62335382 | 685.62338083 | -0.00002701 | 1.0000 |
| 31 | 12 | 30 | 12 | 685.62735712 | 685.62739850 | -0.00004138 | 1.0000 |
| 31 | 13 | 30 | 13 | 685.63173810 | 685.63176580 | -0.00002770 | 1.0000 |
| 31 | 14 | 30 | 14 | 685.63649383 | 685.63648278 | 0.00001105 | 1.0000 |
| 31 | 15 | 30 | 15 | 685.64152368 | 685.64154950 | -0.00002582 | 1.0000 |
| 31 | 16 | 30 | 16 | 685.64694547 | 685.64696606 | -0.00002059 | 1.0000 |
| 31 | 17 | 30 | 17 | 685.65277056 | 685.65273251 | 0.00003805 | 1.0000 |
| 31 | 18 | 30 | 18 | 685.65878570 | 685.65884896 | -0.00006326 | 1.0000 |
| 31 | 19 | 30 | 19 | 685.66538049 | 685.66531548 | 0.00006501 | 1.0000 |
| 31 | 20 | 30 | 20 | 685.67211609 | 685.67213217 | -0.00001608 | 1.0000 |
| 31 | 21 | 30 | 21 | 685.67930278 | 685.67929913 | 0.00000365 | 1.0000 |
| 31 | 22 | 30 | 22 | 685.68681136 | 685.68681647 | -0.00000511 | 1.0000 |
| 31 | 23 | 30 | 23 | 685.69469539 | 685.69468430 | 0.00001109 | 1.0000 |
| 31 | 24 | 30 | 24 | 685.70290394 | 685.70290272 | 0.00000122 | 1.0000 |
| 31 | 25 | 30 | 25 | 685.71144448 | 685.71147186 | -0.00002738 | 1.0000 |
| 31 | 26 | 30 | 26 | 685.72051144 | 685.72039185 | 0.00011959 | 1.0000 |
| 31 | 27 | 30 | 27 | 685.72968624 | 685.72966281 | 0.00002343 | 1.0000 |
| 31 | 28 | 30 | 28 | 685.73930334 | 685.73928488 | 0.00001846 | 1.0000 |
| 31 | 29 | 30 | 29 | 685.74937200 | 685.74925820 | 0.00011380 | 0.0100 |
| 31 | 30 | 30 | 30 | 685.75989200 | 685.75958291 | 0.00030909 | 0.0100 |
| 32 | 0 | 31 | 0 | 685.97304900 | 685.97267275 | 0.00037625 | 0.0033 |
| 32 | 2 | 31 | 2 | 685.97304900 | 685.97337172 | -0.00032272 | 0.0033 |
| 32 | 1 | 31 | 1 | 685.97304900 | 685.97284749 | 0.00020151 | 0.0033 |
| 32 | 3 | 31 | 3 | 685.97405800 | 685.97424544 | -0.00018744 | 0.0050 |
| 32 | 4 | 31 | 4 | 685.97541500 | 685.97546867 | -0.00005367 | 0.0100 |
| 32 | 5 | 31 | 5 | 685.97704106 | 685.97704142 | -0.00000036 | 1.0000 |
| 32 | 6 | 31 | 6 | 685.97897253 | 685.97896372 | 0.00000881 | 1.0000 |
| 32 | 7 | 31 | 7 | 685.98122534 | 685.98123560 | -0.00001026 | 1.0000 |
| 32 | 8 | 31 | 8 | 685.98389983 | 685.98385709 | 0.00004274 | 1.0000 |
| 32 | 9 | 31 | 9 | 685.98682955 | 685.98682822 | 0.00000133 | 1.0000 |
| 32 | 10 | 31 | 10 | 685.99017607 | 685.99014904 | 0.00002703 | 1.0000 |
| 32 | 11 | 31 | 11 | 685.99377717 | 685.99381960 | -0.00004243 | 1.0000 |
| 32 | 12 | 31 | 12 | 685.99788524 | 685.99783996 | 0.00004528 | 1.0000 |
| 32 | 13 | 31 | 13 | 686.00224352 | 686.00221016 | 0.00003336 | 1.0000 |
| 32 | 14 | 31 | 14 | 686.00689521 | 686.00693028 | -0.00003507 | 1.0000 |
| 32 | 15 | 31 | 15 | 686.01198625 | 686.01200038 | -0.00001413 | 1.0000 |
| 32 | 16 | 31 | 16 | 686.01745153 | 686.01742054 | 0.00003099 | 1.0000 |
| 32 | 17 | 31 | 17 | 686.02317637 | 686.02319084 | -0.00001447 | 1.0000 |
| 32 | 18 | 31 | 18 | 686.02938322 | 686.02931136 | 0.00007186 | 1.0000 |
| 32 | 19 | 31 | 19 | 686.03576794 | 686.03578218 | -0.00001424 | 1.0000 |
| 32 | 20 | 31 | 20 | 686.04261210 | 686.04260341 | 0.00000869 | 1.0000 |
| 32 | 21 | 31 | 21 | 686.04977516 | 686.04977515 | 0.00000001 | 1.0000 |
| 32 | 22 | 31 | 22 | 686.05730327 | 686.05729749 | 0.00000578 | 1.0000 |
| 32 | 23 | 31 | 23 | 686.06516469 | 686.06517055 | -0.00000586 | 1.0000 |
| 32 | 24 | 31 | 24 | 686.07337607 | 686.07339444 | -0.00001837 | 1.0000 |
| 32 | 25 | 31 | 25 | 686.08199092 | 686.08196928 | 0.00002164 | 1.0000 |
| 32 | 26 | 31 | 26 | 686.09084711 | 686.09089520 | -0.00004809 | 1.0000 |
| 32 | 27 | 31 | 27 | 686.10016690 | 686.10017233 | -0.00000543 | 1.0000 |
| 32 | 28 | 31 | 28 | 686.10982099 | 686.10980080 | 0.00002019 | 1.0000 |
| 32 | 29 | 31 | 29 | 686.11982966 | 686.11978075 | 0.00004891 | 1.0000 |
| 32 | 30 | 31 | 30 | 686.13017933 | 686.13011233 | 0.00006700 | 1.0000 |
| 32 | 31 | 31 | 31 | 686.14070949 | 686.14079569 | -0.00008620 | 1.0000 |
| 33 | 0 | 32 | 0 | 686.34305500 | 686.34279786 | 0.00025714 | 0.0033 |
| 33 | 2 | 32 | 2 | 686.34305500 | 686.34349729 | -0.00044229 | 0.0033 |
| 33 | 1 | 32 | 1 | 686.34305500 | 686.34297272 | 0.00008228 | 0.0033 |
| 33 | 3 | 32 | 3 | 686.34414200 | 686.34437160 | -0.00022960 | 0.0100 |
| 33 | 4 | 32 | 4 | 686.34551700 | 686.34559564 | -0.00007864 | 0.0100 |
| 33 | 5 | 32 | 5 | 686.34718501 | 686.34716943 | 0.00001558 | 1.0000 |
| 33 | 6 | 32 | 6 | 686.34909299 | 686.34909301 | -0.00000002 | 1.0000 |
| 33 | 7 | 32 | 7 | 686.35136498 | 686.35136639 | -0.00000141 | 1.0000 |
| 33 | 8 | 32 | 8 | 686.35385363 | 686.35398962 | -0.00013599 | 1.0000 |
| 33 | 9 | 32 | 9 | 686.35688900 | 686.35696272 | -0.00007372 | 0.0100 |
| 33 | 10 | 32 | 10 | 686.36031580 | 686.36028574 | 0.00003006 | 1.0000 |
| 33 | 11 | 32 | 11 | 686.36393551 | 686.36395874 | -0.00002323 | 1.0000 |
| 33 | 12 | 32 | 12 | 686.36796766 | 686.36798176 | -0.00001410 | 1.0000 |
| 33 | 13 | 32 | 13 | 686.37231624 | 686.37235486 | -0.00003862 | 1.0000 |
| 33 | 14 | 32 | 14 | 686.37708622 | 686.37707811 | 0.00000811 | 1.0000 |
| 33 | 15 | 32 | 15 | 686.38218077 | 686.38215157 | 0.00002920 | 1.0000 |
| 33 | 16 | 32 | 16 | 686.38755819 | 686.38757532 | -0.00001713 | 1.0000 |
| 33 | 17 | 32 | 17 | 686.39333330 | 686.39334944 | -0.00001614 | 1.0000 |
| 33 | 18 | 32 | 18 | 686.39947626 | 686.39947402 | 0.00000224 | 1.0000 |
| 33 | 19 | 32 | 19 | 686.40593403 | 686.40594913 | -0.00001510 | 1.0000 |
| 33 | 20 | 32 | 20 | 686.41278245 | 686.41277488 | 0.00000757 | 1.0000 |
| 33 | 21 | 32 | 21 | 686.41996361 | 686.41995137 | 0.00001224 | 1.0000 |
| 33 | 22 | 32 | 22 | 686.42748240 | 686.42747869 | 0.00000371 | 1.0000 |
| 33 | 23 | 32 | 23 | 686.43535667 | 686.43535697 | -0.00000030 | 1.0000 |
| 33 | 24 | 32 | 24 | 686.44364563 | 686.44358630 | 0.00005933 | 1.0000 |
| 33 | 25 | 32 | 25 | 686.45215699 | 686.45216683 | -0.00000984 | 1.0000 |
| 33 | 26 | 32 | 26 | 686.46110912 | 686.46109866 | 0.00001046 | 1.0000 |
| 33 | 27 | 32 | 27 | 686.47034836 | 686.47038193 | -0.00003357 | 1.0000 |
| 33 | 28 | 32 | 28 | 686.48013000 | 686.48001677 | 0.00011323 | 0.0100 |
| 33 | 29 | 32 | 29 | 686.49002391 | 686.49000333 | 0.00002058 | 1.0000 |
| 33 | 30 | 32 | 30 | 686.50021603 | 686.50034175 | -0.00012572 | 1.0000 |
| 33 | 31 | 32 | 31 | 686.51088100 | 686.51103217 | -0.00015117 | 0.0100 |
| 34 | 0 | 33 | 0 | 686.71294800 | 686.71262280 | 0.00032520 | 0.0017 |
| 34 | 1 | 33 | 1 | 686.71294800 | 686.71279777 | 0.00015023 | 0.0017 |
| 34 | 2 | 33 | 2 | 686.71294800 | 686.71332269 | -0.00037469 | 0.0017 |
| 34 | 3 | 33 | 3 | 686.71416100 | 686.71419757 | -0.00003657 | 0.0100 |
| 34 | 4 | 33 | 4 | 686.71537500 | 686.71542242 | -0.00004742 | 0.0100 |
| 34 | 5 | 33 | 5 | 686.71702766 | 686.71699726 | 0.00003040 | 1.0000 |
| 34 | 6 | 33 | 6 | 686.71893055 | 686.71892210 | 0.00000845 | 1.0000 |
| 34 | 7 | 33 | 7 | 686.72115787 | 686.72119698 | -0.00003911 | 1.0000 |
| 34 | 8 | 33 | 8 | 686.72386562 | 686.72382194 | 0.00004368 | 1.0000 |
| 34 | 9 | 33 | 9 | 686.72685511 | 686.72679700 | 0.00005811 | 1.0000 |
| 34 | 10 | 33 | 10 | 686.73008765 | 686.73012222 | -0.00003457 | 1.0000 |
| 34 | 11 | 33 | 11 | 686.73382593 | 686.73379764 | 0.00002829 | 1.0000 |
| 34 | 12 | 33 | 12 | 686.73781232 | 686.73782331 | -0.00001099 | 1.0000 |
| 34 | 13 | 33 | 13 | 686.74217537 | 686.74219930 | -0.00002393 | 1.0000 |
| 34 | 14 | 33 | 14 | 686.74694023 | 686.74692567 | 0.00001456 | 1.0000 |
| 34 | 15 | 33 | 15 | 686.75204746 | 686.75200248 | 0.00004498 | 1.0000 |
| 34 | 16 | 33 | 16 | 686.75741419 | 686.75742981 | -0.00001562 | 1.0000 |
| 34 | 17 | 33 | 17 | 686.76323972 | 686.76320774 | 0.00003198 | 1.0000 |
| 34 | 18 | 33 | 18 | 686.76926933 | 686.76933635 | -0.00006702 | 1.0000 |
| 34 | 19 | 33 | 19 | 686.77581327 | 686.77581573 | -0.00000246 | 1.0000 |
| 34 | 20 | 33 | 20 | 686.78264484 | 686.78264599 | -0.00000115 | 1.0000 |
| 34 | 21 | 33 | 21 | 686.78982357 | 686.78982720 | -0.00000363 | 1.0000 |
| 34 | 22 | 33 | 22 | 686.79735764 | 686.79735949 | -0.00000185 | 1.0000 |
| 34 | 23 | 33 | 23 | 686.80529493 | 686.80524296 | 0.00005197 | 1.0000 |
| 34 | 24 | 33 | 24 | 686.81348336 | 686.81347772 | 0.00000564 | 1.0000 |
| 34 | 25 | 33 | 25 | 686.82203609 | 686.82206390 | -0.00002781 | 1.0000 |
| 34 | 26 | 33 | 26 | 686.83098067 | 686.83100162 | -0.00002095 | 1.0000 |
| 34 | 27 | 33 | 27 | 686.84033575 | 686.84029101 | 0.00004474 | 1.0000 |
| 34 | 28 | 33 | 28 | 686.85001914 | 686.84993220 | 0.00008694 | 1.0000 |
| 34 | 29 | 33 | 29 | 686.85980743 | 686.85992533 | -0.00011790 | 1.0000 |
| 34 | 30 | 33 | 30 | 686.87020391 | 686.87027056 | -0.00006665 | 1.0000 |
| 34 | 31 | 33 | 31 | 686.88095800 | 686.88096803 | -0.00001003 | 0.0100 |
| 34 | 32 | 33 | 32 | 686.89206120 | 686.89201789 | 0.00004331 | 1.0000 |
| 34 | 33 | 33 | 33 | 686.90338598 | 686.90342031 | -0.00003433 | 1.0000 |
| 35 | 0 | 34 | 0 | 687.08245200 | 687.08214702 | 0.00030498 | 0.0033 |
| 35 | 1 | 34 | 1 | 687.08245200 | 687.08232211 | 0.00012989 | 0.0033 |
| 35 | 2 | 34 | 2 | 687.08245200 | 687.08284738 | -0.00039538 | 0.0033 |
| 35 | 3 | 34 | 3 | 687.08357400 | 687.08372283 | -0.00014883 | 0.0100 |
| 35 | 4 | 34 | 4 | 687.08497700 | 687.08494848 | 0.00002852 | 0.0100 |
| 35 | 5 | 34 | 5 | 687.08655233 | 687.08652435 | 0.00002798 | 1.0000 |
| 35 | 6 | 34 | 6 | 687.08846779 | 687.08845046 | 0.00001733 | 1.0000 |
| 35 | 7 | 34 | 7 | 687.09078726 | 687.09072684 | 0.00006042 | 1.0000 |
| 35 | 8 | 34 | 8 | 687.09338836 | 687.09335352 | 0.00003484 | 1.0000 |
| 35 | 9 | 34 | 9 | 687.09632175 | 687.09633054 | -0.00000879 | 1.0000 |
| 35 | 10 | 34 | 10 | 687.09967056 | 687.09965794 | 0.00001262 | 1.0000 |
| 35 | 11 | 34 | 11 | 687.10336036 | 687.10333578 | 0.00002458 | 1.0000 |
| 35 | 12 | 34 | 12 | 687.10735008 | 687.10736409 | -0.00001401 | 1.0000 |
| 35 | 13 | 34 | 13 | 687.11175168 | 687.11174295 | 0.00000873 | 1.0000 |
| 35 | 14 | 34 | 14 | 687.11649440 | 687.11647242 | 0.00002198 | 1.0000 |
| 35 | 15 | 34 | 15 | 687.12153213 | 687.12155257 | -0.00002044 | 1.0000 |
| 35 | 16 | 34 | 16 | 687.12701930 | 687.12698346 | 0.00003584 | 1.0000 |
| 35 | 17 | 34 | 17 | 687.13272800 | 687.13276518 | -0.00003718 | 1.0000 |
| 35 | 18 | 34 | 18 | 687.13891236 | 687.13889782 | 0.00001454 | 1.0000 |
| 35 | 19 | 34 | 19 | 687.14539629 | 687.14538146 | 0.00001483 | 1.0000 |
| 35 | 20 | 34 | 20 | 687.15222388 | 687.15221619 | 0.00000769 | 1.0000 |
| 35 | 21 | 34 | 21 | 687.15939766 | 687.15940212 | -0.00000446 | 1.0000 |
| 35 | 22 | 34 | 22 | 687.16698559 | 687.16693935 | 0.00004624 | 1.0000 |
| 35 | 23 | 34 | 23 | 687.17481227 | 687.17482799 | -0.00001572 | 1.0000 |
| 35 | 24 | 34 | 24 | 687.18304698 | 687.18306816 | -0.00002118 | 1.0000 |
| 35 | 25 | 34 | 25 | 687.19168835 | 687.19165997 | 0.00002838 | 1.0000 |
| 35 | 26 | 34 | 26 | 687.20068928 | 687.20060355 | 0.00008573 | 1.0000 |
| 35 | 27 | 34 | 27 | 687.20990100 | 687.20989903 | 0.00000197 | 1.0000 |
| 35 | 28 | 34 | 28 | 687.21951419 | 687.21954654 | -0.00003235 | 1.0000 |
| 35 | 29 | 34 | 29 | 687.22948409 | 687.22954623 | -0.00006214 | 1.0000 |
| 35 | 30 | 34 | 30 | 687.23992399 | 687.23989824 | 0.00002575 | 1.0000 |
| 35 | 31 | 34 | 31 | 687.25059868 | 687.25060272 | -0.00000404 | 1.0000 |
| 35 | 32 | 34 | 32 | 687.26176378 | 687.26165983 | 0.00010395 | 1.0000 |
| 35 | 33 | 34 | 33 | 687.27309382 | 687.27306972 | 0.00002410 | 1.0000 |
| 35 | 34 | 34 | 34 | 687.28502400 | 687.28483256 | 0.00019144 | 0.0100 |
| 36 | 0 | 35 | 0 | 687.45178600 | 687.45137006 | 0.00041594 | 0.0017 |
| 36 | 1 | 35 | 1 | 687.45178600 | 687.45154527 | 0.00024073 | 0.0017 |
| 36 | 2 | 35 | 2 | 687.45178600 | 687.45207088 | -0.00028488 | 0.0017 |
| 36 | 3 | 35 | 3 | 687.45286200 | 687.45294690 | -0.00008490 | 0.0100 |
| 36 | 4 | 35 | 4 | 687.45423182 | 687.45417336 | 0.00005846 | 1.0000 |
| 36 | 5 | 35 | 5 | 687.45577431 | 687.45575026 | 0.00002405 | 1.0000 |
| 36 | 6 | 35 | 6 | 687.45767689 | 687.45767763 | -0.00000074 | 1.0000 |
| 36 | 7 | 35 | 7 | 687.45999595 | 687.45995549 | 0.00004046 | 1.0000 |
| 36 | 8 | 35 | 8 | 687.46268688 | 687.46258389 | 0.00010299 | 1.0000 |
| 36 | 9 | 35 | 9 | 687.46551447 | 687.46556285 | -0.00004838 | 1.0000 |
| 36 | 10 | 35 | 10 | 687.46890785 | 687.46889243 | 0.00001542 | 1.0000 |
| 36 | 11 | 35 | 11 | 687.47257307 | 687.47257267 | 0.00000040 | 1.0000 |
| 36 | 12 | 35 | 12 | 687.47660686 | 687.47660362 | 0.00000324 | 1.0000 |
| 36 | 13 | 35 | 13 | 687.48102032 | 687.48098534 | 0.00003498 | 1.0000 |
| 36 | 14 | 35 | 14 | 687.48569939 | 687.48571790 | -0.00001851 | 1.0000 |
| 36 | 15 | 35 | 15 | 687.49081996 | 687.49080136 | 0.00001860 | 1.0000 |
| 36 | 16 | 35 | 16 | 687.49619571 | 687.49623581 | -0.00004010 | 1.0000 |
| 36 | 17 | 35 | 17 | 687.50203980 | 687.50202131 | 0.00001849 | 1.0000 |
| 36 | 18 | 35 | 18 | 687.50815496 | 687.50815795 | -0.00000299 | 1.0000 |
| 36 | 19 | 35 | 19 | 687.51464088 | 687.51464582 | -0.00000494 | 1.0000 |
| 36 | 20 | 35 | 20 | 687.52149229 | 687.52148502 | 0.00000727 | 1.0000 |
| 36 | 21 | 35 | 21 | 687.52869294 | 687.52867565 | 0.00001729 | 1.0000 |
| 36 | 22 | 35 | 22 | 687.53621063 | 687.53621780 | -0.00000717 | 1.0000 |
| 36 | 23 | 35 | 23 | 687.54412416 | 687.54411160 | 0.00001256 | 1.0000 |
| 36 | 24 | 35 | 24 | 687.55233476 | 687.55235714 | -0.00002238 | 1.0000 |
| 36 | 25 | 35 | 25 | 687.56100640 | 687.56095456 | 0.00005184 | 1.0000 |
| 36 | 26 | 35 | 26 | 687.56988462 | 687.56990398 | -0.00001936 | 1.0000 |
| 36 | 27 | 35 | 27 | 687.57919399 | 687.57920553 | -0.00001154 | 1.0000 |
| 36 | 28 | 35 | 28 | 687.58891415 | 687.58885934 | 0.00005481 | 1.0000 |
| 36 | 29 | 35 | 29 | 687.59875139 | 687.59886555 | -0.00011416 | 1.0000 |
| 36 | 30 | 35 | 30 | 687.60924723 | 687.60922432 | 0.00002291 | 1.0000 |
| 36 | 31 | 35 | 31 | 687.61995986 | 687.61993578 | 0.00002408 | 1.0000 |
| 36 | 32 | 35 | 32 | 687.63101409 | 687.63100010 | 0.00001399 | 1.0000 |
| 36 | 33 | 35 | 33 | 687.64242901 | 687.64241743 | 0.00001158 | 1.0000 |
| 36 | 34 | 35 | 34 | 687.65417651 | 687.65418795 | -0.00001144 | 1.0000 |
| 36 | 35 | 35 | 35 | 687.66620600 | 687.66631181 | -0.00010581 | 0.0100 |
| 37 | 0 | 36 | 0 | 687.82062200 | 687.82029151 | 0.00033049 | 0.0033 |
| 37 | 2 | 36 | 2 | 687.82062200 | 687.82099278 | -0.00037078 | 0.0033 |
| 37 | 1 | 36 | 1 | 687.82062200 | 687.82046683 | 0.00015517 | 0.0033 |
| 37 | 3 | 36 | 3 | 687.82178100 | 687.82186938 | -0.00008838 | 0.0100 |
| 37 | 4 | 36 | 4 | 687.82305900 | 687.82309663 | -0.00003763 | 0.0100 |
| 37 | 5 | 36 | 5 | 687.82468937 | 687.82467455 | 0.00001482 | 1.0000 |
| 37 | 6 | 36 | 6 | 687.82658267 | 687.82660318 | -0.00002051 | 1.0000 |
| 37 | 7 | 36 | 7 | 687.82877643 | 687.82888253 | -0.00010610 | 1.0000 |
| 37 | 8 | 36 | 8 | 687.83151757 | 687.83151263 | 0.00000494 | 1.0000 |
| 37 | 9 | 36 | 9 | 687.83450808 | 687.83449354 | 0.00001454 | 1.0000 |
| 37 | 10 | 36 | 10 | 687.83783936 | 687.83782528 | 0.00001408 | 1.0000 |
| 37 | 11 | 36 | 11 | 687.84146564 | 687.84150791 | -0.00004227 | 1.0000 |
| 37 | 12 | 36 | 12 | 687.84557061 | 687.84554148 | 0.00002913 | 1.0000 |
| 37 | 13 | 36 | 13 | 687.84990495 | 687.84992606 | -0.00002111 | 1.0000 |
| 37 | 14 | 36 | 14 | 687.85470030 | 687.85466169 | 0.00003861 | 1.0000 |
| 37 | 15 | 36 | 15 | 687.85972042 | 687.85974846 | -0.00002804 | 1.0000 |
| 37 | 16 | 36 | 16 | 687.86520199 | 687.86518644 | 0.00001555 | 1.0000 |
| 37 | 17 | 36 | 17 | 687.87098855 | 687.87097570 | 0.00001285 | 1.0000 |
| 37 | 18 | 36 | 18 | 687.87709100 | 687.87711634 | -0.00002534 | 1.0000 |
| 37 | 19 | 36 | 19 | 687.88360477 | 687.88360843 | -0.00000366 | 1.0000 |
| 37 | 20 | 36 | 20 | 687.89051817 | 687.89045208 | 0.00006609 | 1.0000 |
| 37 | 21 | 36 | 21 | 687.89764865 | 687.89764737 | 0.00000128 | 1.0000 |
| 37 | 22 | 36 | 22 | 687.90513087 | 687.90519443 | -0.00006356 | 1.0000 |
| 37 | 23 | 36 | 23 | 687.91304807 | 687.91309335 | -0.00004528 | 1.0000 |
| 37 | 24 | 36 | 24 | 687.92137868 | 687.92134426 | 0.00003442 | 1.0000 |
| 37 | 25 | 36 | 25 | 687.92991065 | 687.92994727 | -0.00003662 | 1.0000 |
| 37 | 26 | 36 | 26 | 687.93889038 | 687.93890250 | -0.00001212 | 1.0000 |
| 37 | 27 | 36 | 27 | 687.94822905 | 687.94821009 | 0.00001896 | 1.0000 |
| 37 | 28 | 36 | 28 | 687.95788820 | 687.95787017 | 0.00001803 | 1.0000 |
| 37 | 29 | 36 | 29 | 687.96787903 | 687.96788288 | -0.00000385 | 1.0000 |
| 37 | 30 | 36 | 30 | 687.97824841 | 687.97824837 | 0.00000004 | 1.0000 |
| 37 | 31 | 36 | 31 | 687.98894876 | 687.98896679 | -0.00001803 | 1.0000 |
| 37 | 32 | 36 | 32 | 687.99999672 | 688.00003829 | -0.00004157 | 1.0000 |
| 37 | 33 | 36 | 33 | 688.01148285 | 688.01146304 | 0.00001981 | 1.0000 |
| 37 | 34 | 36 | 34 | 688.02322546 | 688.02324119 | -0.00001573 | 1.0000 |
| 37 | 35 | 36 | 35 | 688.03540887 | 688.03537292 | 0.00003595 | 1.0000 |
| 37 | 36 | 36 | 36 | 688.04742500 | 688.04785841 | -0.00043341 | 0.0100 |
| 38 | 0 | 37 | 0 | 688.18926100 | 688.18891103 | 0.00034997 | 0.0033 |
| 38 | 2 | 37 | 2 | 688.18926100 | 688.18961275 | -0.00035175 | 0.0033 |
| 38 | 1 | 37 | 1 | 688.18926100 | 688.18908646 | 0.00017454 | 0.0033 |
| 38 | 3 | 37 | 3 | 688.19035600 | 688.19048991 | -0.00013391 | 0.0100 |
| 38 | 4 | 37 | 4 | 688.19165700 | 688.19171796 | -0.00006096 | 0.0100 |
| 38 | 5 | 37 | 5 | 688.19330435 | 688.19329691 | 0.00000744 | 1.0000 |
| 38 | 6 | 37 | 6 | 688.19518728 | 688.19522678 | -0.00003950 | 1.0000 |
| 38 | 7 | 37 | 7 | 688.19755833 | 688.19750761 | 0.00005072 | 1.0000 |

| 38 | 8  | 37 | 8  | 688.20011022 | 688.20013942 | -0.00002920 | 1.0000 |
|----|----|----|----|--------------|--------------|-------------|--------|
| 38 | 9  | 37 | 9  | 688.20313545 | 688.20312225 | 0.00001320  | 1.0000 |
| 38 | 10 | 37 | 10 | 688.20643230 | 688.20645615 | -0.00002385 | 1.0000 |
| 38 | 11 | 37 | 11 | 688.21015445 | 688.21014116 | 0.00001329  | 1.0000 |
| 38 | 12 | 37 | 12 | 688.21417774 | 688.21417735 | 0.00000039  | 1.0000 |
| 38 | 13 | 37 | 13 | 688.21852909 | 688.21856476 | -0.00003567 | 1.0000 |
| 38 | 14 | 37 | 14 | 688.22329289 | 688.22330346 | -0.00001057 | 1.0000 |
| 38 | 15 | 37 | 15 | 688.22838865 | 688.22839353 | -0.00000488 | 1.0000 |
| 38 | 16 | 37 | 16 | 688.23379500 | 688.23383502 | -0.00004002 | 0.0100 |
| 38 | 17 | 37 | 17 | 688.23963429 | 688.23962803 | 0.00000626  | 1.0000 |
| 38 | 18 | 37 | 18 | 688.24575338 | 688.24577264 | -0.00001926 | 1.0000 |
| 38 | 19 | 37 | 19 | 688.25226200 | 688.25226893 | -0.00000693 | 0.0050 |
| 38 | 20 | 37 | 20 | 688.25907036 | 688.25911701 | -0.00004665 | 1.0000 |
| 38 | 21 | 37 | 21 | 688.26628676 | 688.26631696 | -0.00003020 | 1.0000 |
| 38 | 22 | 37 | 22 | 688.27388953 | 688.27386890 | 0.00002063  | 1.0000 |
| 38 | 23 | 37 | 23 | 688.28181100 | 688.28177293 | 0.00003807  | 1.0000 |
| 38 | 24 | 37 | 24 | 688.29000168 | 688.29002918 | -0.00002750 | 1.0000 |
| 38 | 25 | 37 | 25 | 688.29859842 | 688.29863775 | -0.00003933 | 1.0000 |
| 38 | 26 | 37 | 26 | 688.30763146 | 688.30759877 | 0.00003269  | 1.0000 |
| 38 | 27 | 37 | 27 | 688.31690136 | 688.31691238 | -0.00001102 | 1.0000 |
| 38 | 28 | 37 | 28 | 688.32658524 | 688.32657870 | 0.00000654  | 1.0000 |
| 38 | 29 | 37 | 29 | 688.33655631 | 688.33659789 | -0.00004158 | 1.0000 |
| 38 | 30 | 37 | 30 | 688.34698471 | 688.34697008 | 0.00001463  | 1.0000 |